\documentclass[journal,10pt]{IEEEtran}
\usepackage{amssymb}
\usepackage{amsmath}
\usepackage{epsfig}
\usepackage{epstopdf}
\usepackage{cite}
\usepackage{bm}
\usepackage[noend]{algpseudocode}
\usepackage{lettrine}
\usepackage{algorithmicx,algorithm}
\usepackage{color}
\usepackage{subfigure}
\usepackage{balance}
\usepackage{enumitem}
\IEEEoverridecommandlockouts

\graphicspath{ {figures/}}

\begin{document}
	
	\title{Optimal Interference Exploitation Waveform Design with Relaxed Block-Level Power Constraints}

	\author{Xiao Tong,~\IEEEmembership{Graduate~Student~Member,~IEEE}, Lei Lei,~\IEEEmembership{Senior~Member,~IEEE}, Ang Li,~\IEEEmembership{Senior~Member,~IEEE},\\ A. Lee Swindlehurst,~\IEEEmembership{Fellow,~IEEE} and Symeon Chatzinotas,~\IEEEmembership{Fellow,~IEEE}
		\thanks{
			X. Tong, L. Lei and A. Li are with the School of Information and Communications Engineering, Faculty of Electronic and Information Engineering, Xi’an Jiaotong University, Xi’an, Shaanxi 710049, China (e-mail: xiao.tong.2023@stu.xjtu.edu.cn; lei.lei@xjtu.edu.cn; ang.li.2020@xjtu.edu.cn).
			
			A. Lee Swindlehurst is with the Center for Pervasive Communications and Computing, University of California at Irvine, Irvine, CA 92697 USA (e-mail: swindle@uci.edu).
			
			S. Chatzinotas is with Interdisciplinary Center for Security, Reliability and Trust (SnT), University of Luxembourg, 4365 Esch-sur-Alzette, Luxembourg (e-mail: symeon.chatzinotas@uni.lu).
		}
	}
	
	\markboth{}
	{}
	
	\maketitle
	
	\begin{abstract}
		This paper investigates constructive interference (CI)-based waveform design for phase shift keying and quadrature amplitude modulation symbols under relaxed block-level power constraints in multi-user multiple-input single-output (MU-MIMO) communication systems. Existing linear CI-based precoding methods, including symbol-level precoding (SLP) and block-level precoding (BLP), suffer from performance limitations due to strict symbol-level power budgets or insufficient degrees of freedom (DoFs) over the block. To overcome these challenges, we propose a nonlinear waveform optimization framework that introduces additional optimization variables and maximizes the minimum CI metric across the transmission block. The optimal waveform is derived in closed form using the Lagrangian function and Karush–Kuhn–Tucker conditions, and the solution is explicitly expressed with respect to the dual variables. Moreover, the original problems are equivalently reformulated as tractable quadratic programming (QP) problems. To efficiently solve the derived QP problems, we develop an improved alternating direction method of multipliers (ADMM) algorithm by integrating a linear-time projection technique, which significantly enhances the computational efficiency. Simulation results demonstrate that the proposed algorithms substantially outperform the conventional CI-SLP and CI-BLP approaches, particularly under high-order modulations and large block lengths.
	\end{abstract}
	
	
	\IEEEpeerreviewmaketitle
	
	\section{Introduction}\label{introduction}
	\lettrine[lines=2]{M}{ultiple}-input multiple-output (MIMO) technology is a cornerstone of modern wireless communications, offering significant improvements in spectral efficiency and system capacity by exploiting spatial multiplexing and diversity \cite{mm MIMO, Large MIMO}. However, as the number of users increases, multi-user interference (MUI) becomes a major challenge that can degrade system performance \cite{MIMO Detection}. To address this challenge, various precoding techniques have been developed to suppress interference by leveraging channel state information (CSI), enabling MIMO systems to fully utilize the available spatial degrees of freedom (DoFs) and enhance reliability and throughput \cite{MIMO Linear Precoding}.
	
	MIMO precoding methods are typically categorized into linear and nonlinear approaches. In linear precoding, a matrix is designed to linearly combine the transmit symbols in an effort to suppress interference. Classical approaches like matched filtering (MF) \cite{MF}, zero-forcing (ZF) \cite{ZF}, and regularized ZF (RZF) \cite{RZF} use CSI to compute closed-form precoders. However, these approaches often suffer from degraded performance in scenarios with low signal-to-noise ratio (SNR) or highly correlated channels. To overcome these issues, optimization-based linear precoding has emerged as a promising alternative, aiming to maximize signal-to-interference-plus-noise ratio (SINR) \cite{SINR balancing} or minimize transmit power \cite{Power Minimization}. Nevertheless, these problems are usually non-convex due to nonlinear objectives or constraints. Techniques like semidefinite programming (SDP) and semidefinite relaxation (SDR) are commonly used to obtain convex approximations \cite{SDR-star,SDR-ISAC}. Yet, the removal of rank-one constraints during relaxation often leads to suboptimal solutions \cite{SDR}. A more effective strategy is to directly optimize the transmit waveform, bypassing precoder design. This paradigm, which generally results in nonlinear precoding, enables more precise control over the transmit signal and has demonstrated superior performance in various scenarios \cite{Fan-ISAC,Rang-ISAC}.
	
	Unlike traditional interference suppression methods, interference exploitation precoding introduces the concept of constructive interference (CI), where multi-user interference (MUI) is leveraged to enhance signal detection \cite{Ang CI survey,Symeon survey}. CI precoding deliberately steers interference toward constructive regions by jointly exploiting CSI and the known transmit symbols. As a result, the precoding matrix must be updated on a symbol-by-symbol basis, giving rise to the symbol-level precoding (SLP) framework adopted in most CI schemes. In \cite{CF-PSK,CF-QAM}, closed-form SLP solutions for Phase Shift Keying (PSK) and Quadrature Amplitude Modulation (QAM) modulation in multi-user (MU) Multiple-Input Single-Output (MISO) systems are derived using a Lagrangian and Karush–Kuhn–Tucker (KKT) analysis. More general MU-MIMO scenarios are explored in \cite{RIRC,SLPMLD}, where both the transmit precoder and receiver combining are jointly optimized. Specifically, \cite{RIRC} proposes a novel regularized interference rejection combiner (RIRC), while \cite{SLPMLD} integrates CI with maximum likelihood detection (MLD) to improve decoding performance. Owing to these benefits, CI-based SLP has been extended to various applications, including integrated sensing and communication (ISAC) \cite{SLP isac1,SLP isac2}, intelligent reflecting surface (IRS)-assisted systems \cite{SLP IRS1,SLP IRS2}, and physical-layer security \cite{SLP S1,SLP S2}.
	
	While SLP methods offer significant performance gains, the required symbol-by-symbol processing imposes high computational complexity. To address this challenge, \cite{Group SLP} proposes a grouped SLP scheme, where users with similar channels are grouped to exploit intra-group interference while suppressing inter-group interference using conventional techniques. Although this reduces complexity, it retains the symbol-level processing paradigm. In contrast, \cite{BLP CI} introduces a CI-based block-level precoding (CI-BLP) algorithm that applies a fixed precoding matrix to a block of symbols with static CSI, similar to traditional BLP frameworks \cite{ZF,RZF,SINR balancing,Power Minimization,SDR-star,SDR-ISAC,SDR,Fan-ISAC,Rang-ISAC}. Unlike conventional BLP, the CI-BLP matrix depends on both CSI and the user symbols, enabling constructive use of MUI within the block. Although the use of a fixed matrix reduces the number of available DoFs, CI-BLP benefits from relaxed block-level power constraints, which allows more flexible power allocation. As shown in \cite{BLP CI}, the CI-BLP approach outperforms CI-SLP in terms of symbol error rate (SER) when the block length is relatively small. This performance advantage, along with reduced complexity, has also been validated in ISAC scenarios \cite{BLP ISAC}, where CI-BLP achieves superior SER and computational efficiency compared to CI-SLP.
	
	Although CI-BLP reduces computational complexity while achieving satisfactory communication performance, it fails to attain the maximum possible CI gains due to its reliance on linear precoding. The fundamental limitation lies in the fact that, in practical systems, block lengths are typically large—often comprising dozens or even hundreds of symbols. Employing a constant precoding matrix across such a block severely restricts the system’s design flexibility, thereby making it challenging to obtain a near-optimal transmit waveform. To address this challenge, nonlinear waveform design schemes have been proposed to directly construct the transmit waveform and achieve optimal system performance. For example, \cite{Spano and Alodeh}  and \cite{X ISAC} enhance communication performance in MIMO systems by designing the transmit waveform and exploiting CI gains. However, these approaches still perform symbol-by-symbol optimization, which results in high computational complexity.
	
	To address the aforementioned challenges, this paper proposes CI-based waveform designs for PSK and QAM modulation with block-level power budgets in a MU-MISO downlink communication system. The proposed algorithms further demonstrate the potential of CI techniques in modern wireless communication systems. The main contributions of this work are summarized as follows:
	\begin{enumerate}
		\item We propose optimal CI-based waveform designs for both PSK and QAM modulation applied over a block of transmit symbols. The optimization objective is to maximize the minimum CI metric while satisfying relaxed block-level power constraints. By designing the waveform over a transmission block, our approaches achieve the maximum possible CI performance improvement.
		\item For PSK modulation, we derive a closed-form solution for the optimal transmit waveform using the Lagrangian method and KKT conditions. The original problem is equivalently reformulated as a quadratic programming (QP) problem with fewer variables, substantially reducing the computational complexity.
		\item For QAM modulation, we adopt a generalized ``symbol-scaling'' CI metric and extend the analytical framework developed for the PSK case. The resulting optimization problem is similarly transformed into a QP framework with fewer variables, enabling both computational efficiency and optimal CI performance.
		\item To further improve computational efficiency, we develop an alternating direction method of multipliers (ADMM)-based algorithm capable of efficiently solving the QP problems arising from both the PSK and QAM waveform designs. Within each iteration, closed-form solutions are derived, and a linear-time projection technique is applied to efficiently compute the dual variables.
		\item Our analysis reveals that, compared to conventional CI-SLP schemes, the proposed approaches benefit from relaxed power constraints, thereby enhancing power allocation flexibility. Compared to traditional CI-BLP methods, our nonlinear waveform design over the entire block significantly increases the number of  degrees of design freedom and the CI gains, especially in scenarios with large block lengths and high-order modulations.
	\end{enumerate}
	
	Simulation results validate the effectiveness of the proposed QP formulation and ADMM algorithm in efficiently solving the optimization problems. The proposed waveform design algorithms consistently outperform conventional CI-SLP and CI-BLP methods for various scenarios, with particularly notable SER performance gains achieved for larger block lengths and higher modulation orders.
	
	\subsection{Organization and Notations}\label{OAN}
	The remainder of the paper is organized as follows: Section \ref{SMCI} provides a brief introduction to the system model and the CI constraints. Section \ref{ATCIBS} reviews the conventional CI-SLP and CI-BLP approaches. In Section \ref{POCIWDPSK}, the optimal waveform design scheme for PSK modulation is developed in closed form, and an ADMM algorithm is proposed to solve the resulting QP problem. Section \ref{POCIWDQAM} extends the waveform design to QAM modulation and derives the corresponding closed-form transmit waveform. Section \ref{Simulation} presents simulation results to validate the performance advantages and computational efficiency of the proposed algorithms for various scenarios. Finally, Section \ref{Conclusion} concludes the paper and discusses potential future research directions.
	
	Notations: Bold uppercase and lowercase letters represent matrices and vectors, respectively. The symbols $\mathbb{C}^{M \times N}$ and $\mathbb{R}^{M \times N}$ denote space of complex and real-valued matrices of size $M \times N$, while $\mathbb{C}^{M \times 1}$ and $\mathbb{R}^{M \times 1}$ denote complex and real-valued vectors of length $M$. For a matrix $\mathbf{A}$, we use $\mathbf{A}^T$, $\mathbf{A}^*$, $\mathbf{A}^H$, $\mathbf{A}^{-1}$, and $\mathrm{Tr}(\mathbf{A})$ to denote its transpose, conjugate, Hermitian transpose, inverse, and trace, respectively. The operators $\mathcal{R}(\cdot)$ and $\mathcal{J}(\cdot)$ extract the real and imaginary parts of a complex number. The norms $||\cdot||_F$ and $||\cdot||_2$ denote the Frobenius norm for matrices and the Euclidean norm for vectors, respectively. The cardinality of a set is denoted by $card(\cdot)$. The operator $\mathrm{diag}(\mathbf{a})$ forms a diagonal matrix with the entries of vector $\mathbf{a}$ on its main diagonal, while $\mathrm{blkdiag}(\mathbf{A}_1, \dots, \mathbf{A}_n)$ constructs a block diagonal matrix with $\mathbf{A}_1, \dots, \mathbf{A}_n$ along its diagonal blocks.
	
	\section{System Model and Constructive Interference}\label{SMCI}
	
	\subsection{System Model}\label{SysMod}
	\begin{figure}[t]
		\centering 
		\subfigure[CI-SLP]{
			\label{CI-SLP}
			\includegraphics[scale=0.3]{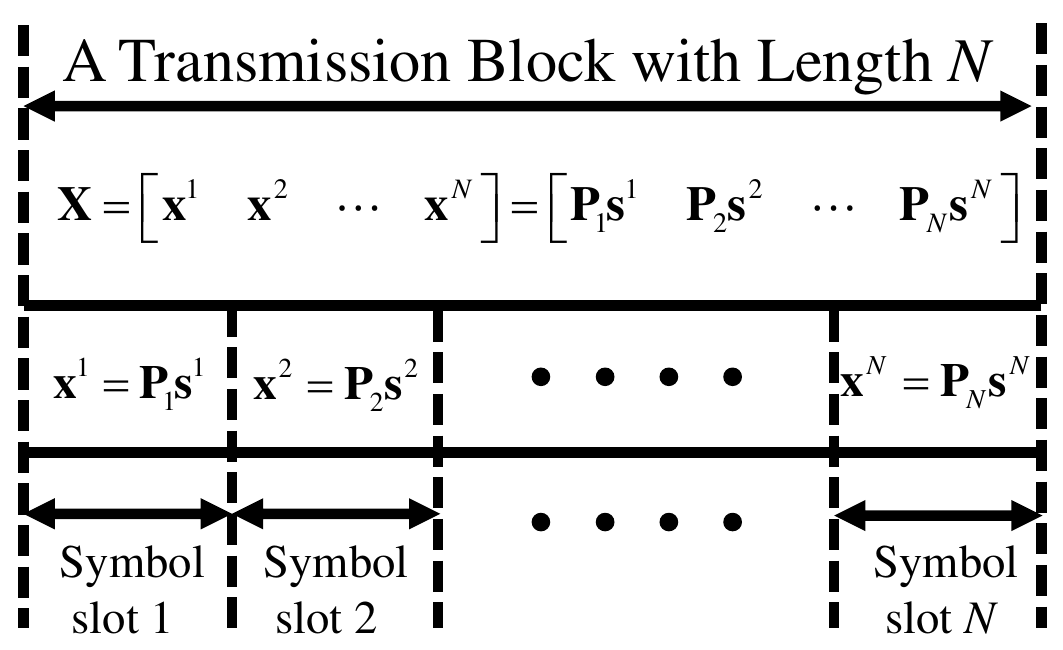}}
		\subfigure[CI-BLP]{
			\label{CI-BLP}
			\includegraphics[scale=0.3]{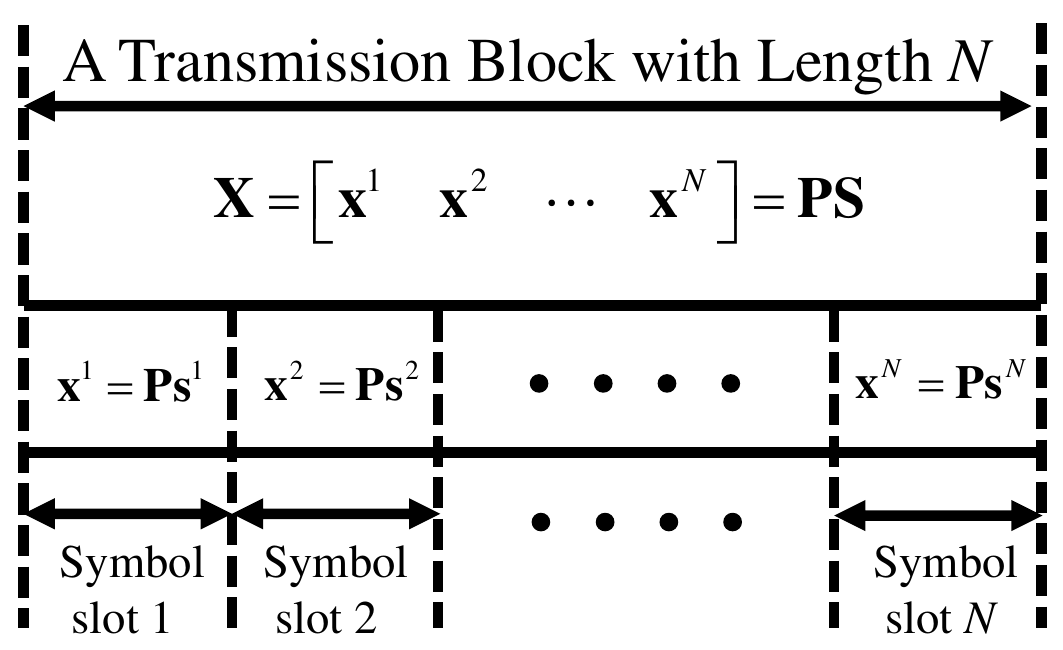}}
		\caption{Difference between CI-SLP and CI-BLP schemes.}
		\label{Method}
	\end{figure}
	We consider a MU-MISO downlink communication system, where a BS with $N_T$ antennas transmits independent data streams to $K \le N_T$ users. We define ${{\bf{s}}^n} = {\left[ {s_1^n,s_2^n, \cdots ,s_K^n} \right]^T} \in \mathbb{C} {^{K \times 1}}$ as the transmit symbol vector in the \textit{n}-th symbol slot, and there are $N$ symbol slots in the considered block, which is assumed to be smaller than the channel coherence interval. Therefore, the transmit symbol matrix for the block can be expressed as
	\begin{equation}
		{\bf{S}} = \left[ {{{\bf{s}}^1},{{\bf{s}}^2}, \cdots ,{{\bf{s}}^N}} \right] \in \mathbb{C} {^{K \times N}}.
	\end{equation}
	The transmit symbol vector is multiplied by a precoding matrix before transmission. The precoding matrix can be categorized as either symbol-level or block-level, depending on whether it varies across different transmit symbol vectors in different symbol slots. For SLP, the received signal for the \textit{k}-th user in the \textit{n}-th slot is represented by \cite{CF-PSK,CF-QAM}
	\begin{equation}\label{SLP y}
		y_k^n = {\bf{h}}_k^T{{\bf{P}}_n}{{\bf{s}}^n} + z_k^n,\forall k \in {\cal K},\forall n \in {\cal N},
	\end{equation}
	where ${{\bf{h}}_k} \in \mathbb{C} {^{{N_T} \times 1}}$ is the channel vector between the BS and the \textit{k}-th user, assumed to be constant during a transmission block, and $z_k^n$ represents Gaussian noise with zero mean and variance $\sigma^2$ at the receiver. The matrix ${{\bf{P}}_n} \in \mathbb{C} {^{{N_T} \times K}}$ denotes the SLP matrix for the transmit symbol vector in the \textit{n}-th slot, which varies for different symbol slots. On the other hand, the BLP matrix remains unchanged within a block, and the received signal for the \textit{k}-th user in the \textit{n}-th slot can be expressed as \cite{BLP CI}
	\begin{equation}\label{BLP y}
		y_k^n = {\bf{h}}_k^T{{\bf{P}}}{{\bf{s}}^n} + z_k^n,\forall k \in {\cal K},\forall n \in {\cal N},
	\end{equation}
	where ${{\bf{P}}} \in \mathbb{C} {^{{N_T} \times K}}$ represents the BLP matrix, which is used for all symbol slots in the block.
	
	Fig. \ref{Method} highlights the fundamental distinction between CI-SLP and CI-BLP: CI-SLP designs a unique precoding matrix for each symbol slot, offering high communication performance but limited by symbol-level power constraints. In contrast, CI-BLP employs a fixed precoding matrix across the entire symbol block, which substantially reduces computational complexity but sacrifices design flexibility and adaptability.
	
	\subsection{Constructive Interference}\label{CI}
	
	\begin{figure}[t]
		\centering 
		\includegraphics[width=2.2in]{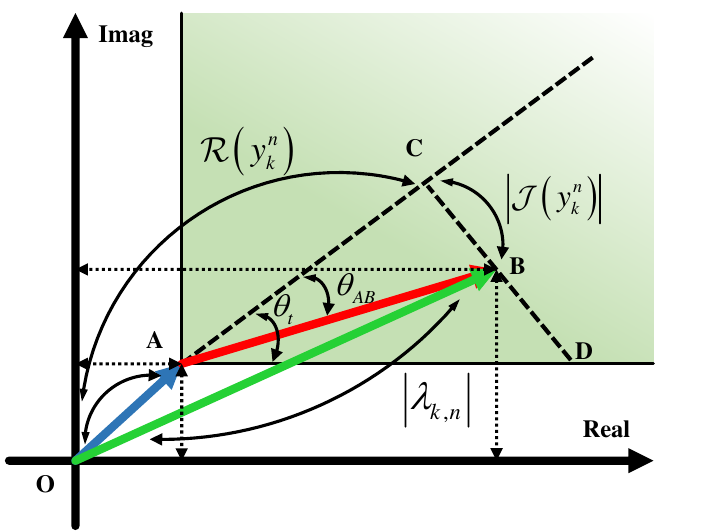}
		\caption{Constructive interference, QPSK.}
		\label{PSK}
	\end{figure}
	
	To facilitate the understanding of CI, Fig. \ref{PSK} illustrates the CI region in the first quadrant for QPSK modulated symbols. In the figure, $\overrightarrow {OB} = {\bf{h}}_k^T{{\bf{P}}}{{\bf{s}}^n} ={\lambda _{k,n}}{s^n_{k}} $ represents the noiseless received signal for the \textit{k}-th user in the \textit{n}-th slot, and ${\lambda _{k,n}}$ can be complex, which represents a transformation that applies both a phase rotation and amplitude scaling to the original symbol ${s_{k,n}}$. The vector $\overrightarrow {OA} = t \cdot {s^n_{k}}$ denotes a scaled version of the symbol ${s^n_{k}}$. The scaling factor $t$ reflects the detection threshold: a larger value of $t$ implies a longer distance from the decision boundaries, which is beneficial for signal detection. Therefore, CI-based schemes typically aim to maximize $t$ for a given transmit power level. As shown in the figure, to ensure that the received signal falls within the CI region (the green area), $\tan {\theta _{AB}} \le \tan {\theta _t}$ must be satisfied, which leads to the following formulation of the CI constraint for PSK modulation \cite{CF-PSK,RIRC}:
	\begin{equation}\label{PSK CI constraint}
		\begin{array}{l}
			{\kern 12pt} \tan {\theta _{AB}} \le \tan {\theta _t}  \vspace{1ex}\\
			\Rightarrow \left[ {{\cal R}\left( {{\lambda _{k,n}}} \right) - t} \right]\tan {\theta _t} \ge |{\cal I}\left( {{\lambda _{k,n}}} \right)|,
		\end{array}
	\end{equation}
	where ${\theta _t}= \dfrac{\pi }{\mathcal{M}}$ is the threshold angle for $\mathcal{M}$-PSK modulation.
	
	\section{Fundamental Limitations of Traditional CI-SLP and CI-BLP} \label{ATCIBS}
	Although existing precoding methods based on CI have demonstrated the ability to enhance the useful signal strength by leveraging both CSI and transmit data information, their performance bottlenecks in block transmission scenarios remain insufficiently understood and analyzed. Consequently, current CI-SLP  and CI-BLP approaches struggle to fully exploit the potential CI gain. To address this issue, we first analyze the fundamental limitations of traditional CI-SLP and CI-BLP schemes and then proposes an enhanced precoding design that achieves the performance limit of CI-based waveform design.
	
	\subsection{Traditional CI-SLP}\label{Tcislp}
	The original CI-SLP approach for PSK modulation was introduced in \cite{CF-PSK}, where the minimum distance between the received signal and the decision boundaries is maximized to enhance the CI effect, subject to constructive interference constraints and symbol-level power budgets. The corresponding optimization problem is formulated based on (\ref{SLP y}) and (\ref{PSK CI constraint}), and can be expressed as \cite{CF-PSK}
	\begin{equation}\label{P1}
		\begin{array}{l}
			{\mathcal{P}}_1:{\kern 1pt} \mathop {\max }\limits_{{{\bf{P}}_n}}{\kern 2pt} t \vspace{1ex}\\
			{\kern 27pt} s.t.{\kern 4pt} {\bf{h}}_k^T{{\bf{P}}_n}{{\bf{s}}^n} = {\lambda _{k,n}}s_k^n,\forall k \in {\cal K},\forall n \in {\cal N} \vspace{1ex}\\
			{\kern 42pt} \left[ {{\cal R}\left( {{\lambda _{k,n}}} \right) - t} \right] \tan {\theta _t} \ge \left| {{\cal I}\left( {{\lambda _{k,n}}} \right)} \right| \vspace{1ex}\\
			{\kern 42pt} \left\| {{{\bf{P}}_n}{{\bf{s}}^n}} \right\|_2^2 \le p_0,\forall n \in {\cal N},
		\end{array}
	\end{equation}
	where ${{{\bf{P}}_n}}$ is the SLP matrix corresponding to the data vector ${{\bf{s}}^n}$. The optimal solution for ${\mathcal P}_1$ can be expressed as \cite{CF-PSK}
	\begin{equation}\label{PSK closed solution}
		{{\bf{P}}_n} = \frac{1}{K}{{\bf{H}}^H}{\left( {{\bf{H}}{{\bf{H}}^H}} \right)^{ - 1}}diag\left\{ {\sqrt {\dfrac{{{p_0}}}{{{\bf{u}}_u^T{\bf{V}}_n^{ - 1}{{\bf{u}}_n}}}} {\bf{V}}_n^{ - 1}{{\bf{u}}_n}} \right\}{{\bf{s}}^n}{{\bf{\hat s}}^n},
	\end{equation}
	where 
	\begin{subequations}
		\begin{align}
			{{\bf{H}}} &= \left[ {{\bf h}_1},{{\bf h}_2}, \cdots ,{{\bf h}_K} \right]^T \in \mathbb{C} {^{K \times N_T}}, \vspace{1ex}\\ 
			{{\bf{\hat s}}^n} &= \left[ {\frac{1}{{s_1^n}},\frac{1}{{s_2^n}}, \cdots ,\frac{1}{{s_K^n}}} \right] \in \mathbb{C} {^{1 \times K}}, \vspace{1ex}\\  
			{{\bf{u}}_n} &= {\left[ {{\mu _{n,1}},{\mu _{n,2}}, \cdots ,{\mu _{n,K}}} \right]^T}\in \mathbb{R} {^{K \times 1}},  \vspace{1ex}\\ 
			{{\bf{V}}_n} &= {\mathcal{R}}\left[ {diag\left( {({{{\bf{ s}}}^n})}^H \right){{\left( {{\bf{H}}{{\bf{H}}^H}} \right)}^{ - 1}}diag\left( {{{\bf{s}}^n}} \right)} \right],
		\end{align}
	\end{subequations}
	and ${{\bf{u}}_n}$ represents the column dual variable vector for the \textit{n}-th symbol slot sub-problem, which can be obtained by solving the following QP problem:
	\begin{equation}\label{P2}
		\begin{array}{l}
			{\mathcal{P}}_2:{\kern 1pt} \mathop {\min }\limits_{{{\bf{u}}_n}} {\kern 1pt} {\kern 1pt}{\bf{u}}_n^T{\bf{V}}_n^{ - 1}{{\bf{u}}_n} \vspace{1ex}\\
			{\kern 25pt}s.t.{\kern 4pt}  {{\bf{1}}^T}{{\bf{u}}_n} = 1 \vspace{1ex}\\
			{\kern 44pt}{\mu _{k,n}} \ge 0,\forall k \in {\cal K}.
		\end{array}
	\end{equation}
	Details can be found in \cite{CF-PSK}.
	
	Through the above derivation, the original problem $\mathcal{P}_1$, which involves $N_T \times K$ variables, is equivalently transformed into a QP problem $\mathcal{P}_2$ with only $K$ variables, and the overall computational complexity is substantially reduced. Moreover, it can be observed from the formulations of both $\mathcal{P}_1$ and $\mathcal{P}_2$ that the SLP matrix can be independently optimized for each symbol slot. This inter-slot independence stems from the symbol-level power constraints, which enforce a fixed transmit power per slot and thus preclude dynamic power allocation across the block. While this constraint has limited impact when the block length is small, it becomes a major performance bottleneck in large block length scenarios, significantly restricting the achievable CI gain of conventional CI-SLP algorithms.
	
	\subsection{Traditional CI-BLP}\label{Tciblp}
	The CI-BLP approach, first proposed in \cite{BLP CI}, uses a fixed precoding matrix over the entire transmission block, significantly reducing the number of optimization variables and complexity compared to CI-SLP, which requires solving $N$ separate problems. However, this simplification reduces design flexibility. When the block length $N$ is small, sufficient DoFs remain to achieve notable CI gains. As $N$ grows much larger than the number of transmit antennas, the limited number of variables restricts the performance. The block-level power constraint in CI-BLP enables flexible power allocation, which can partially offset this degradation, allowing CI-BLP to outperform CI-SLP when $N$ is not too large.
	
	In the design of CI-BLP the ``symbol-scaling'' CI metric is used to decompose the signal into two real-valued parameters. The received noiseless signal can be expressed as
	\begin{equation}
		{\bf{h}}_k^T{{\bf{P}}}{{\bf{s}}^n} = \tau _{k,{\mathcal{A}}}^ns_{k,{\mathcal{A}}}^n + \tau _{k,{\mathcal{B}}}^ns_{k,{\mathcal{B}}}^n,
	\end{equation}
	where $s_{k,{\mathcal{A}}}^n$ and $s_{k,{\mathcal{B}}}^n $ are the two vector components of ${\bf s}^n_k$ along the detection threshold directions, and $\alpha _{k,{\mathcal{A}}}^n$ and $\alpha _{k,{\mathcal{B}}}^n$ are two real-valued scaling parameters. The CI-BLP problem can be formulated as \cite{BLP CI}
	\begin{equation}\label{P3}
		\begin{array}{l}
			{\mathcal{P}}_3:{\kern 1pt} {\kern 1pt} {\kern 1pt} \mathop {\max }\limits_{{{\bf{P}}_\text{E}}} {\kern 1pt} {\kern 1pt} {\kern 1pt} {\kern 1pt} \mathop {\min {\kern 1pt} }\limits_{k,n} {\kern 1pt} {\kern 1pt} \tau _k^n \vspace{1ex}\\
			{\kern 28pt}s.t.{\kern 8pt} {\kern 1pt} {\bm{\tau }}_\text{E}^n = {{\bf{M}}^n}{{\bf{P}}_\text{E}}{\bf{s}}_\text{E}^n,\forall n \in {\cal N} \vspace{1ex}\\
			{\kern 50pt}\sum\limits_{n = 1}^N {\left\| {{{\bf{P}}_\text{E}}{\bf{s}}_\text{E}^n} \right\|_2^2}  \le N{p_0},
		\end{array}
	\end{equation}
	where 
	\begin{subequations}
		\begin{align}
			{\bm{\tau }}_\text{E}^n &= {\left[ {\tau _{1,{\mathcal{A}}}^n, \cdots ,\tau _{K,{\mathcal{A}}}^n,\tau _{1,{\mathcal{B}}}^n, \cdots ,\tau _{K,{\mathcal{B}}}^n} \right]^T}\in {{\mathbb{R}}^{2K \times 1}},  \vspace{1ex}\\ 
			{{\bf{P}}_\text{E}} &= \left[ {\begin{array}{*{20}{c}}
					{{\mathcal{R}}\left( {\bf{P}} \right)}&{ - {\mathcal{I}}\left( {\bf{P}} \right)}\\
					{{\mathcal{I}}\left( {\bf{P}} \right)}&{{\mathcal{R}}\left( {\bf{P}} \right)}
			\end{array}} \right] \in {{\mathbb{R}}^{2{N_T} \times 2K}}, \vspace{1ex}\\  
			{\bf{s}}_\text{E}^n &= {\left[ {\begin{array}{*{20}{c}}
						{{\mathbb{R}}{{\left( {{{\bf{s}}^n}} \right)}^T}}&{{\mathcal{I}}{{\left( {{{\bf{s}}^n}} \right)}^T}}
				\end{array}} \right]^T}\in {{\mathcal{R}}^{2K \times 1}},  \vspace{1ex}\\ 
			{{\bf{M}}^n} &= {\left[ {{{\bf{j}}_1},{{\bf{j}}_2}, \cdots ,{{\bf{j}}_K},{{\bf{l}}_1},{{\bf{l}}_2}, \cdots ,{{\bf{l}}_K}} \right]^T}\in {{\mathbb{R}}^{2K \times 2{N_T}}},
		\end{align}
	\end{subequations}
	and ${{\bf{j}}_k}\in {{\mathbb{R}}^{2{N_T} \times 1}}$ and ${{\bf{l}}_k}\in {{\mathbb{R}}^{2{N_T} \times 1}}$ can be found in (38) and (39) in \cite{BLP CI}. To solve ${\mathcal{P}}_3$, the variables are transformed into real-valued representations, i.e., ${\bf{\hat P}} = \left[ {\begin{array}{*{20}{c}}
			{{\mathcal{R}}\left( {\bf{P}} \right)}&{{\mathcal{I}}\left( {\bf{P}} \right)}
	\end{array}} \right] \in {\mathbb{R}^{{N_T} \times 2K}}$. Based on this transformation, the optimal closed-form structure for ${\bf{\hat P}}$ can be expressed as
	\begin{equation}
		{\bf{\hat P}} = \frac{1}{{2{\hat\nu} }}\sum\limits_{n = 1}^N {\left[ {{{\left( {{{\bf{A}}^n}} \right)}^T}{{\bm{\delta }}^n}{{\left( {{\bf{s}}_{\text{E}}^n} \right)}^T} + {{\left( {{{\bf{B}}^n}} \right)}^T}{{\bm{\delta }}^n}{{\left( {{\bf{c}}_{\text{E}}^n} \right)}^T}} \right]} {{\bf{D}}^{ - 1}},
	\end{equation}
	where ${\hat \nu}$ is the non-negative Lagrange parameter related to the block-level power budget, and the other variables are given by
	\begin{subequations}
		\begin{align}
			{{\bf{A}}^n} &= {{\bf{M}}^n}{\left[ {{{\bf{I}}_{{N_T}}},{{\bf{0}}_{{N_T}}}} \right]^T} \in {{\mathbb{R}}^{2{N_T} \times {N_T}}}, \vspace{1ex}\\ 
			{{\bf{B}}^n} &= {{\bf{M}}^n}{\left[ {{{\bf{0}}_{{N_T}}},{{\bf{I}}_{{N_T}}}} \right]^T} \in {{\mathbb{R}}^{2{N_T} \times {N_T}}}, \vspace{1ex}\\
			{\bf{c}}_E^n &= \left[ {\begin{array}{*{20}{c}}
					{\bf{0}}&{{{\bf{I}}_K}}\\
					{ - {{\bf{I}}_K}}&{\bf{0}}
			\end{array}} \right]{\bf{s}}_E^n \in {{\mathbb{R}}^{2K \times 1}}, \vspace{1ex}\\
			{\bf{D}} &= \sum\limits_{n = 1}^N {{\bf{s}}_{\text{E}}^n{{\left( {{\bf{s}}_{\text{E}}^n} \right)}^T} + } \sum\limits_{n = 1}^N {{\bf{c}}_{\text{E}}^n{{\left( {{\bf{c}}_{\text{E}}^n} \right)}^T}}\in {{\mathbb{R}}^{2K \times 2K}},
		\end{align}
	\end{subequations}
	and ${{\bm{\delta }}^n} = \left[ {\delta _1^n,\delta _1^n, \cdots ,\delta _{2K}^n} \right]^T \in {{\mathbb{R}}^{2K \times 1}}$ is the Lagrange vector associated with the first constraint in ${\mathcal{P}}_3$. We further combine the $N$ Lagrange vectors together as
	\begin{equation}
		{{\bm{\delta }}_{\text{E}}} = {\left[ {{{\left( {{{\bm{\delta }}^n}} \right)}^T},{{\left( {{{\bm{\delta }}^n}} \right)}^T}, \cdots ,{{\left( {{{\bm{\delta }}^N}} \right)}^T}} \right]^T} \in {\mathbb{R}^{2NK \times 1}},
	\end{equation}
	which can be obtained by solving the following QP problem:
	\begin{equation}\label{P4}
		\begin{array}{l}
			{\mathcal{P}}_4:{\kern 2pt} \mathop {\min }\limits_{{{\bm{\delta }}_E}} {\kern 1pt} {\bm{\delta }}_E^T{\bf{U}}{{\bm{\delta }}_E} \vspace{1ex}\\
			{\kern 25pt}s.t.{\kern 2pt} {\kern 1pt} {{\bf{1}}^T}{{\bm{\delta }}_E} - 1 = 0, \vspace{1ex}\\
			{\kern 42pt}\delta _E^i \ge 0,\forall i \in \left\{ {1,2, \cdots ,2KN} \right\},
		\end{array}
	\end{equation}
	where ${\bf{U}}$ is an intermediate term associated with the transmit data symbols. Its explicit form, along with the detailed derivation from $\mathcal{P}_3$ to $\mathcal{P}_4$, can be found in\cite{BLP CI}, and is omitted here.
	
	Based on the above derivation, the original CI-BLP problem $\mathcal{P}_3$ is equivalently reformulated as a QP problem $\mathcal{P}_4$, where the optimization variables correspond to the dual variables of the $NK$ transmit symbols, resulting in a real-valued dimension of $2NK$, which is larger than the dimension of the ${N_T} \times K$ precoding matrix in $\mathcal{P}_3$. Although $\mathcal{P}_4$ offers a closed-form solution structure, its higher variable dimension may lead to increased complexity when using conventional solvers, especially in practical systems where $N \gg N_T$. In comparison, CI-SLP reduces the original problem $\mathcal{P}_1$ to $N$ decoupled QP problems $\mathcal{P}_2$, each with $K$ variables, totaling $NK$ paramters. The key distinction lies in the constraint structure: CI-SLP uses symbol-level power constraints, while CI-BLP's block-level formulation in $\mathcal{P}_4$ introduces coupling among the variables, allowing for more flexible power allocation and greater optimization freedom.
	
	\section{Proposed CI Waveform Design for PSK Modulation} \label{POCIWDPSK}
	Due to the performance loss incurred by using a fixed precoding matrix in CI-BLP and the rigid symbol-level power constraint in CI-SLP, both methods fail to achieve the maximum possible CI gain, which motivates the nonlinear waveform design proposed in this section.
	
	\subsection{Problem Formulation}
	We propose enhancing the symbol-level precoding framework by introducing a block-level power constraint, which effectively addresses the limitations of CI-SLP and CI-BLP and achieves the maximum possible CI gain. The problem can be formulated as
	\begin{equation}\label{P5}
		\begin{array}{l}
			{\mathcal{P}}_5:\mathop {\max }\limits_{{{\bf{P}}_n}} {\kern 1pt} {\kern 1pt} {\kern 1pt} {\kern 1pt} t \vspace{1ex}\\
			{\kern 23pt} s.t.{\kern 7pt} {\bf{h}}_k^T{{\bf{P}}_n}{{\bf{s}}^n} = {\lambda _{k,n}}s_k^n,\forall k \in {\cal K},\forall n \in {\cal N} \vspace{1ex}\\
			{\kern 43pt} \left[ {{\cal R}\left( {{\lambda _{k,n}}} \right) - t} \right]\tan {\theta _t} \ge \left| {{\cal I}\left( {{\lambda _{k,n}}} \right)} \right| \vspace{1ex}\\
			{\kern 43pt} \left\| {{\bf{P\bar S}}} \right\|_F^2 \le N{p_0},
		\end{array}
	\end{equation}
	where 
	\begin{subequations}
		\begin{align}
			{\bf{P}} &= \left[ {{{\bf{P}}_1},{{\bf{P}}_2}, \cdots ,{{\bf{P}}_N}} \right] \in \mathbb{C} {^{{N_T} \times NK}}, \vspace{1ex}\\ 
			\bar{\mathbf{S}} &= \operatorname{blkdiag}(\mathbf{s}^1, \mathbf{s}^2, \ldots, \mathbf{s}^N) \in \mathbb{C}^{NK \times N}.
		\end{align}
	\end{subequations}
	Compared to ${\mathcal{P}}_1$, ${\mathcal{P}}_5$ introduces a block-level power constraint that enhances power allocation flexibility and increases the number of available DoFs. Unlike ${\mathcal{P}}_3$, ${\mathcal{P}}_5$ optimizes an SLP matrix of dimension $N_T \times NK$, significantly expanding the variable space and improving CI gain. However, while the optimal transmit waveform can be obtained via the linear SLP method in (\ref{P5}), it incurs prohibitively high computational complexity. Conversely, block-level linear precoding falls short of optimality due to its limited number of DoFs. To overcome these limitations, we propose a more compact and direct nonlinear CI-based waveform design, formulated as follows:
	\begin{equation}\label{P6}
		\begin{array}{l}
			{\mathcal{P}}_6:\mathop {\max }\limits_{{{\bf{X}}}} {\kern 1pt} {\kern 1pt} {\kern 1pt} {\kern 1pt} t \vspace{1ex}\\
			{\kern 23pt} s.t.{\kern 7pt}  {\bf{H}}{{\bf{X}}} = {\bf{\Lambda}}{\bf{\bar S}} \vspace{1ex}\\
			{\kern 42pt} \left[ {{\cal R}\left( {{\lambda _{k,n}}} \right) - t} \right]\tan {\theta _t} \ge \left| {{\cal I}\left( {{\lambda _{k,n}}} \right)} \right| \vspace{1ex}\\
			{\kern 43pt}  {\left\| {{{\bf{X}}}} \right\|_F^2}  \le N{p_0},
		\end{array}
	\end{equation}
	where ${\bf{X}} \in \mathbb{C} {^{{N_T} \times N}}$ represents the transmit waveform, and $\mathbf{\Lambda} = \mathrm{blkdiag}\left[ \mathrm{diag}(\bm{\lambda}_1), \mathrm{diag}(\bm{\lambda}_2), \ldots, \mathrm{diag}(\bm{\lambda}_N)\right]\in \mathbb{C}^{K \times KN}$denotes a complex scaling factor matrix. Compared to ${\mathcal{P}}_1$, ${\mathcal{P}}_6$ introduces a relaxed block-level power constraint, thereby enhancing power allocation flexibility. Compared to ${\mathcal{P}}_3$, ${\mathcal{P}}_6$ not only offers a higher dimensional variable space and thus greater DoFs, but also overcomes the fundamental limitation of the BLP matrix in ${\mathcal{P}}_3$, which is incapable of achieving the optimal transmit waveform. Compared to ${\mathcal{P}}_5$, ${\mathcal{P}}_6$ directly designs the transmit waveform rather than the SLP matrix, which substantially decreases the number of optimization variables and significantly reduces the computational complexity. The following proposition establishes the equivalence between ${\mathcal{P}}_5$ and ${\mathcal{P}}_6$:
	
	\textit{Proposition 1}: \({\mathcal{P}}_5\) and \({\mathcal{P}}_6\) are equivalent, as they achieve the same optimal objective value, and their optimal solutions correspond bijectively via the transformation $\mathbf{X}^{\star} = \mathbf{P}^{\star} \bar{\mathbf{S}}$.
	
	\textit{Proof}: See Appendix A.
	
	\subsection{Closed-form Solution Structure of $\mathbf{X}$}
	Although ${\mathcal{P}}_6$ is already a convex second-order cone programming (SOCP) problem and can be solved using standard tools, we obtain the optimal closed-form structure for ${\bf X}$ to reduce the computational complexity. To this end, we employ the Lagrange and KKT methods to analyze ${\mathcal{P}}_6$ and derive its optimal closed-form solution. Specifically, ${\mathcal{P}}_6$ can be rewritten in a more compact form as
	\begin{equation}
		\begin{array}{l}
			{\mathcal{P}}_6:\mathop {\max }\limits_{{{\bf{x}}^n}} {\kern 2pt} t \vspace{1ex}\\
			{\kern 23pt} s.t.{\kern 1pt} {\kern 1pt} {\kern 1pt} {\kern 1pt} {\kern 1pt} {\bf{h}}_k^T{{\bf{x}}^n} = {{\lambda} _{k,n}}s_k^n,\forall k \in {\cal K},\forall n \in {\cal N} \vspace{1ex}\\
			{\kern 42pt}\dfrac{{{\cal I}\left( {{{\lambda} _{k,n}}} \right)}}{{\tan {\theta _t}}} + t - {\cal R}\left( {{{ \lambda} _{k,n}}} \right) \le 0,\forall k \in {\cal K},\forall n \in {\cal N} \vspace{1ex}\\
			{\kern 42pt}- \dfrac{{{\cal I}\left( {{{\lambda} _{k,n}}} \right)}}{{\tan {\theta _t}}} + t - {\cal R}\left( {{{ \lambda} _{k,n}}} \right) \le 0,\forall k \in {\cal K},\forall n \in {\cal N} \vspace{1ex}\\
			{\kern 43pt} \sum\limits_{n = 1}^N {\left\| {{{\bf{x}}^n}} \right\|_2^2}  \le N{p_0},
		\end{array}
	\end{equation}
	where ${\lambda_{k,n}}$ denote the complex scaling factors and the CI constraint is rewritten as two separate constraints. The Lagrangian function of ${\mathcal{P}}_6$ can be expressed as
	\begin{equation}\label{L1}
		\begin{array}{l}
			{\cal L}\left( {{{\bf{x}}^n},t,{\delta _{k,n}},{\alpha _{k,n}},{\beta _{k,n}},{\delta _0}} \right) = \vspace{1ex}\\
			- t + \sum\limits_{n = 1}^N {\sum\limits_{k = 1}^K {{\delta _{k,n}}\left( {{{\bf{h}}_k^T}{{\bf{x}}^n} - {\lambda _{k,n}}{s_{k,n}}} \right)} } \vspace{1ex}\\
			+ \sum\limits_{n = 1}^N {\sum\limits_{k = 1}^K {{\alpha _{k,n}}\left[ {\dfrac{{{\cal I}\left( {{\lambda _{k,n}}} \right)}}{{\tan {\theta _t}}} + t - {\cal R}\left( {{\lambda _{k,n}}} \right)} \right]} } \vspace{1ex}\\
			+ \sum\limits_{n = 1}^N {\sum\limits_{k = 1}^K {{\beta _{k,n}}\left[ { - \dfrac{{{\cal I}\left( {{\lambda _{k,n}}} \right)}}{{\tan {\theta _t}}} + t - {\cal R}\left( {{\lambda _{k,n}}} \right)} \right]} } \vspace{1ex}\\
			 + {\delta _0}\left( {\sum\limits_{n = 1}^N {\left\| {{{\bf{x}}^n}} \right\|_2^2}  - N{p_0}} \right),
		\end{array}
	\end{equation}
	where ${\delta _{k,n}}$, ${\alpha _{k,n}} \ge 0$, ${\beta _{k,n}} \ge 0$ and ${\delta _0} \ge 0$ are the dual variables, and ${\delta _{k,n}}$ may be complex as it is the dual variable related to the equality constraint. 
	
	The KKT conditions for (\ref{L1}) are given by
	\begin{subequations}
		\begin{align}
			\dfrac{{\partial {\cal L}}}{{\partial t}} = {\sum\limits_{n = 1}^N {\sum\limits_{k = 1}^K {{\alpha _{k,n}}} }}+{\sum\limits_{n = 1}^N {\sum\limits_{k = 1}^K {{\beta _{k,n}}} }} - 1 &= 0, \vspace{1ex}\\ 
			\frac{{\partial {\cal L}}}{{\partial {{\bf{x}}^n}}} = \sum\limits_{k = 1}^K {{\delta _{k,n}}{\bf{h}}_k}  + 2\delta _0 {{{\bf{x}}^n}} &= {\bf{0}},\forall n \in {\cal N}, \label{KKT2} \vspace{1ex}\\ 
			{{{\bf{h}}_k^T}{{\bf{x}}^n} - {\lambda _{k,n}}{s_{k,n}}} &= 0,\forall k \in {\cal K},\forall n \in {\cal N}, \label{KKT3} \vspace{1ex}\\
			{\delta _0}\left( {\sum\limits_{n = 1}^N {\left\| {{{\bf{x}}^n}} \right\|_2^2}  - N{p_0}} \right) &= 0,
		\end{align}
	\end{subequations}
	where the KKT conditions corresponding to the inactive CI constraints are omitted for simplicity. Based on (\ref{KKT2}), it can be observed that ${\delta _0} > 0$. Furthermore, ${{\bf{x}}^n}$ can be obtained as
	\begin{equation}\label{xn}
		\begin{array}{l}
			\sum\limits_{k = 1}^K {{\delta _{k,n}}{\bf{h}}_k}  + 2\delta _0 {{{\bf{x}}^n}} = {\bf{0}}\\
			\Rightarrow {{{\bf{x}}^n}} =  - \dfrac{1}{{2\delta _0 }}\sum\limits_{k = 1}^K {{\delta _{k,n}}{\bf{h}}_k} \\
			\Rightarrow {{{\bf{x}}^n}}  = \sum\limits_{k = 1}^K {\varsigma _k^n{\bf{h}}_k}  = {{\bf{H}}^T}{{\bm{\varsigma }}^n},\\
		\end{array}
	\end{equation}
		where we define $\varsigma _k^n \buildrel \Delta \over =  - \dfrac{{{\delta _{k,n}}}}{{2{\delta _0}}}$ and ${{\bm{\varsigma }}^n} = {\left[ {\varsigma _1^n,\varsigma _2^n, \cdots ,\varsigma _K^n} \right]^T} \in \mathbb{C} {^{K \times 1}}$. Then, with each ${{\bf{x}}^n}$ obtained, the waveform $\bf X$ can be expressed as
	\begin{equation}
		\begin{aligned}
			{\bf{X}} &= \left[ {{\bf{x}}^1, {\bf{x}}^2, \cdots, {\bf{x}}^N} \right] \vspace{1ex} \\
			&= \left[ {{\bf{H}}^T {\bm{\varsigma}}^1, {\bf{H}}^T {\bm{\varsigma}}^2, \cdots, {\bf{H}}^T {\bm{\varsigma}}^N} \right] \vspace{1ex} \\
			&= {\bf{H}}^T {\bm{\Delta}},
		\end{aligned}
	\end{equation}
	where we define the dual variable matrix ${\bm{\Delta}} = \left[ {{{\bm{\varsigma }}^1},{{\bm{\varsigma }}^2}, \cdots ,{{\bm{\varsigma }}^N}} \right]\in \mathbb{C} {^{K \times N}}$. Furthermore, we can derive that
	\begin{equation}
		\begin{aligned}
			{\bf{HX}} &= {\bf{\Lambda}}{\bf{\bar S}} \vspace{1ex}\\
			&\Rightarrow {\bf{H}}{{\bf{H}}^T}{\bm{\Delta}} = {\bf{\Lambda}}{\bf{\bar S}} \vspace{1ex}\\
			&\Rightarrow {\bm{\Delta}} = {\left( {{\bf{H}}{{\bf{H}}^T}} \right)^{ - 1}}{\bf{\Lambda}}{\bf{\bar S}},
		\end{aligned}
	\end{equation}
	therefore, the transmit waveform can be expressed in a more compact form as
	\begin{equation}\label{opX}
		{\bf{X}} = {{\bf{H}}^T}{\left( {{\bf{H}}{{\bf{H}}^T}} \right)^{ - 1}}{\bf{\Lambda}}{\bf{\bar S}}.
	\end{equation}
	
	Based on above analysis, we know that $\delta _0 > 0$, which means the block-level power constraint is active when optimality is achieved, i.e., 
	\begin{equation}\label{X}
		\begin{aligned}
			\sum\limits_{n = 1}^N {\left\| {{{\bf{x}}^n}} \right\|_2^2}  &= Np_0\\
			&\Rightarrow \sum\limits_{n = 1}^N {\left\| {{\bf{G}}{{\rm{diag}}\left( {{\bm{\lambda }}_n} \right)}{{\bf{s}}^n}} \right\|_2^2}  = Np_0\\
			&\Rightarrow \sum\limits_{n = 1}^N {\left\| {{\bf{G}}{\rm{diag}}\left( {{{\bf{s}}^n}} \right){{\bm{\lambda }}_n}} \right\|_2^2}  = Np_0\\
			&\Rightarrow \sum\limits_{n = 1}^N {\left( {{\bm{\lambda }}_n^T{\bf{\bar G}}_n^H{{{\bf{\bar G}}}_n}{{\bm{\lambda }}_n}} \right)}  = Np_0\\
			&\Rightarrow {{\bm{\lambda }}^T}{{{\bf{\bar G}}}^H}{\bf{\bar G{\bm{\lambda }} }} = Np_0
		\end{aligned}
	\end{equation}
	where ${\bf{G}} = {{\bf{H}}^T}{\left( {{\bf{H}}{{\bf{H}}^T}} \right)^{ - 1}}\in \mathbb{C} {^{N_T \times K}}$, ${\bf{\bar G}}_n = {\bf{G}}{\rm{diag}}\left( {{{\bf{s}}^n}} \right) \in \mathbb{C} {^{N_T \times K}}$, ${\bm{\lambda }} = {\left[ {{\bm{\lambda }}_1^T,{\bm{\lambda }}_2^T, \cdots ,{\bm{\lambda }}_N^T} \right]^T} \in \mathbb{R} {^{NK \times 1}}$ and $\bar{\mathbf{G}} = \operatorname{blkdiag}(\bar{\mathbf{G}}_1, \ldots, \bar{\mathbf{G}}_N) \in \mathbb{C} {^{NN_T \times NK}}$.
	With (\ref{X}), the original problem ${\mathcal{P}}_6$ can be equivalently transformed as
	\begin{equation}\label{P7}
		\begin{array}{l}
			{\mathcal{P}}_{7}:\mathop {\max }\limits_{\bm{\lambda }} {\kern 3pt}  t \vspace{1ex}\\
			{\kern 24pt} s.t.{\kern 5pt}\dfrac{{{\cal I}\left( {{\lambda _{k,n}}} \right)}}{{\tan {\theta _t}}} + t - {\cal R}\left( {{\lambda _{k,n}}} \right) \le 0,\forall k \in {\cal K},\forall n \in {\cal N} \vspace{1ex}\\
			{\kern 42pt}- \dfrac{{{\cal I}\left( {{\lambda _{k,n}}} \right)}}{{\tan {\theta _t}}} + t - {\cal R}\left( {{\lambda _{k,n}}} \right) \le 0,\forall k \in {\cal K},\forall n \in {\cal N} \vspace{1ex}\\
			{\kern 42pt} {{\bm{\lambda }}^T}{{{\bf{G}}}^H}{\bf{G}}{\bm{\lambda }} - Np_0 = 0.
		\end{array}
	\end{equation}
	
	In order to solve (\ref{P7}), we separate the variables into real and imaginary parts:
	\begin{subequations}
		\begin{align}
			{\bm{\hat \lambda }} &= {\left[ {{\cal{R}}\left( {{{\bm{ \lambda }}^T}} \right),{\cal{I}}\left( {{{\bm{\lambda }}^T}} \right)} \right]^T} \in \mathbb{R} {^{2NK \times 1}},\label{lamdahat} \vspace{1ex}\\ 
			{\bf{\hat G}} &= \left[ {\begin{array}{*{20}{c}}
					{{\cal{R}}\left( {{{{\bf{\bar G}}}^H}{\bf{\bar G}}} \right)}&{ - {\cal{I}}\left( {{{{\bf{\bar G}}}^H}{\bf{\bar G}}} \right)}\\
					{{\cal{I}}\left( {{{{\bf{\bar G}}}^H}{\bf{\bar G}}} \right)}&{{\cal{R}}\left( {{{{\bf{\bar G}}}^H}{\bf{\bar G}}} \right)}
			\end{array}} \right]\in \mathbb{R} {^{2NK \times 2NK}}.
		\end{align}
	\end{subequations}
	Therefore, the power budget of (\ref{P7}) becomes
	\begin{equation}\label{Power}
		{{\bm{\hat \lambda }}^T}{\bf{\hat G}}{\bm{\hat \lambda }}- Np_0 = 0.
	\end{equation}
	The Lagrangian function of (\ref{P7}) is constructed as
	\begin{equation}\label{P90}
		\begin{aligned}
			&{\cal L}\left( {{\kern 1pt} {\bm{\hat \lambda }},t,{{ \alpha }_0},{{ \mu }_{k,n}},{{ \nu }_{k,n}}} \right) =  {{\alpha }_0}{{{{\bm{\hat \lambda }}}^T}{\bf{\hat G }}{\bm{\hat \lambda}} - {{\alpha }_0}N{p_0}} \\
			& \left[ {\sum\limits_{n = 1}^N {\sum\limits_{k = 1}^K {\left( {{{ \mu }_{k,n}} + {{ \nu }_{k,n}}} \right)} }  - 1} \right]t - \sum\limits_{n = 1}^N {\sum\limits_{k = 1}^K {\left( {{{ \mu }_{k,n}} + {{ \nu }_{k,n}}} \right){\cal{R}}\left( {{\lambda _{k,n}}} \right)} }\\
			&+ \sum\limits_{n = 1}^N {\sum\limits_{k = 1}^K {\left( {{{ \mu }_{k,n}} - {{ \nu }_{k,n}}} \right)\dfrac{{{\cal{I}}\left( {{\lambda _{k,n}}} \right)}}{{\tan {\theta _t}}}} },
		\end{aligned}
	\end{equation}
	where ${{\alpha }_0}$, ${{\mu }_{k,n}} \ge 0 $, and ${{ \nu }_{k,n}} \ge 0 $ are the dual variables, and we define
	\begin{equation}\label{P91}
		\begin{array}{l}
			{\bf{ u}} = {\left[ { \cdots ,{{\mu }_{k,n}}, \cdots ,{{ \mu }_{K,N}}, \cdots ,{{\nu }_{k,n}}, \cdots ,{{ \nu }_{K,N}}} \right]^T} \in {\mathbb{C}^{2KN \times 1}},\vspace{1ex}\\
			{\bf{M}} = \left[ {\begin{array}{*{20}{c}}
					{{{\bf{I}}_{KN}}}&{ - \dfrac{1}{{\tan {\theta _t}}}{{\bf{I}}_{KN}}}\\
					{{{\bf{I}}_{KN}}}&{\dfrac{1}{{\tan {\theta _t}}}{{\bf{I}}_{KN}}}
			\end{array}} \right] \in {\mathbb{C}^{2KN \times 2KN}}.
		\end{array}
	\end{equation}
	Then, (\ref{P90}) can be simplified as
	\begin{equation}\label{P92}
		{\cal L}\left( {{\kern 1pt} {\bm{\hat \lambda }},t,{{\alpha }_0},{\bf{u}}} \right) = \left( {{{\bf{1}}^T}{\bf{ u}} - 1} \right)t + { \alpha _0}{{\bm{\hat \lambda }}^T}{\bf{\hat G}}{\bm{\hat \lambda }} - {{\bf{u}}^T}{\bf{M}}{\bm{\hat \lambda }} - {\alpha _0}N{p_0}.
	\end{equation}
	The KKT conditions of (\ref{P92}) are expressed as
	\begin{subequations}
		\begin{align}
			\dfrac{{\partial {\cal L}}}{{\partial t}} = {{\bf{1}}^T}{\bf{ u}} - 1 &= 0, \vspace{1ex}\\ 
			\dfrac{{\partial {\cal L}}}{{\partial {\bm{\hat \lambda }}}} = 2{ \alpha _0}{\bf{\hat G}}{\bm{\hat \lambda }} - {\bf{M}}^T{\bf{ u}} &= {\bf{0}}, \label{KKT4} \vspace{1ex}\\ 
			{{{{\bm{\hat \lambda }}}^T}{\bf{\hat G }}{\bm{\hat \lambda}} - N{p_0}} &= 0, \label{KKT5}
		\end{align}
	\end{subequations}
	where the KKT conditions corresponding to the inactive CI constraints are omitted for simplicity. 
	
	An expression for the optimal scaling ${\bm{\hat \lambda }}$ can be derived based on (\ref{KKT4}):
	\begin{equation}\label{lambdahat}
		{\bm{\hat \lambda }} = \frac{1}{{2{{\alpha }_0}}}{{\bf{\hat G}}^{ - 1}}{{\bf{M}}^T}{\bf{ u}}.
	\end{equation}
	Substituting this expression into (\ref{KKT5}), we have
	\begin{equation}
		\begin{array}{l}
			{{{\bm{\hat \lambda }}}^T}{\bf{\hat G}}{\bm{\hat \lambda}} - N{p_0} = 0 \vspace{1ex}\\
			\Rightarrow {\left( {\dfrac{1}{{2{{\alpha }_0}}}{{{\bf{\hat G}}}^{ - 1}}{{\bf{M}}^T}{\bf{ u}}} \right)^T}{\bf{\hat G}}\left( {\dfrac{1}{{2{{\alpha }_0}}}{{{\bf{\hat G}}}^{ - 1}}{{\bf{M}}^T}{\bf{ u}}} \right) = N{p_0} \vspace{1ex}\\
			\Rightarrow {{\alpha }_0} = \sqrt {\dfrac{{{{{\bf{ u}}}^T}{{{\bf{ V}}}}{\bf{ u}}}}{{4N{p_0}}}},
		\end{array}
	\end{equation}
	where 
	\begin{equation}
		{\bf{V}} = {\bf{M}}{{\bf{\hat G}}^{ - 1}}{{\bf{M}}^T}.
	\end{equation}
	It can be observed that Slater's condition \cite{CF-PSK} is satisfied and thus the dual gap is zero. Thus, the corresponding dual problem can be expressed as
	\begin{equation}
		\begin{aligned}
			{\cal U} &= \mathop {\max }\limits_{{\bf{ u}},{\alpha _0}} {\kern 1pt} {\kern 1pt} {\kern 1pt} \mathop {\min }\limits_{{\bm{\hat \lambda }},t} {\kern 1pt} {\kern 1pt} {\kern 1pt} {\cal L}\left( {{\kern 1pt} {\bm{\hat \lambda }},t,{ \alpha _0},{{\mu }_i}} \right) \vspace{1ex}\\
			&= \mathop {\max }\limits_{{\bf{\bar u}},{\alpha _0}} {\kern 1pt} {\kern 1pt} {\kern 1pt} {\alpha _0}{\left( \frac{1}{{2{{\alpha }_0}}}{{\bf{\hat G}}^{ - 1}}{{\bf{M}}^T}{\bf{ u}} \right)^T}{\bf{\hat G}}\left( \frac{1}{{2{{\alpha }_0}}}{{\bf{\hat G}}^{ - 1}}{{\bf{M}}^T}{\bf{ u}} \right) \vspace{1ex}\\
			&- {{{\bf{ u}}}^T}{\bf{M}}\left( \frac{1}{{2{{\alpha }_0}}}{{\bf{\hat G}}^{ - 1}}{{\bf{M}}^T}{\bf{ u}} \right) - {\alpha _0}N{p_0} \vspace{1ex}\\
			&= \mathop {\max }\limits_{{\bf{ u}}} {\kern 1pt} {\kern 1pt} {\kern 1pt}  - \sqrt {N{p_0}{{{\bf{ u}}}^T}{{\bf{ V}}}{\bf{ u}}}.
		\end{aligned}
	\end{equation}
	Therefore, the dual problem is equivalent to the following QP problem
	\begin{equation}\label{P8}
		\begin{array}{l}
			{\mathcal{P}}_{8}:\mathop {\min }\limits_{{\bf{ u}}} {\kern 2pt} {{{\bf{ u}}}^T}{{\bf{V}}}{\bf{ u}} \vspace{1ex}\\
			{\kern 22pt} s.t.{\kern 4pt} {{\bf{1}}^T}{\bf{ u}} = 1 \vspace{1ex}\\
			{\kern 40pt} {{u }_i} \ge 0,\forall i \in \left\{ {1,2, \cdots ,2KN} \right\}.
		\end{array}
	\end{equation}
	
	In the above derivation, $\mathcal{P}_6$ is first reformulated as $\mathcal{P}_7$ with less variables, which is subsequently transformed into the more tractable convex QP problem $\mathcal{P}_8$. Although this two-step transformation simplifies the original problem, in practical scenarios where $N$ is typically large, the resulting QP problem still involves a considerable number of variables, potentially leading to a high computational burden. To further mitigate the resulting complexity, the following subsection presents an efficient ADMM-based algorithm for solving $\mathcal{P}_8$. 

	\subsection{Efficient ADMM Algorithm for Solving ${\mathcal{P}}_8$}
	To efficiently solve the QP problem $\mathcal{P}_8$ derived above, we develop an ADMM-based algorithm that is specifically designed to reduce computational complexity. Compared with general-purpose convex solvers, the proposed method leverages the problem structure to achieve faster convergence and lower cost per iteration, making it more suitable for large block length scenarios.
	
	By introducing an auxiliary variable $\bf z$, $\mathcal{P}_8$ can be further transformed as
	\begin{equation}\label{P9}
		\begin{array}{l}
			{\mathcal{P}}_{9}:\mathop {\min }\limits_{{\bf{u}},{\bf{z}}} {\kern 1pt} {\kern 1pt} {\kern 1pt} {\kern 1pt} {{\bf{u}}^T}{\bf{Vu}} + {\mathbb I}{_{\mathcal C}}\left( {\bf{z}} \right) \vspace{1ex}\\
			{\kern 21pt} {\kern 1pt} s.t.{\kern 1pt} {\kern 1pt} {\kern 1pt} {\kern 1pt} {\kern 1pt} {\kern 1pt} {\kern 1pt} {\bf{u}} - {\bf{z}} = 0,
		\end{array}
	\end{equation}
	where ${\mathbb I}{_{\mathcal C}}\left( {\bf{z}} \right)$ is the indicator function defined as
	\begin{equation}\label{IC}
		{\mathbb I}{_{\mathcal C}}\left( {\bf{z}} \right) = 
		\begin{cases}
			0, & \text{if } {{\bf{z}} \in {\mathcal C}}, \\
			+\infty, & \text{otherwise},
		\end{cases}
	\end{equation}
	and for ${\mathcal{P}}_{9}$ the feasible region ${\mathcal C}$ is
	\begin{equation}\label{C}
		{\mathcal C} = \left\{ {{\bf{z}}{\kern 2pt}|{\kern 2pt}{{\bf{1}}^T}{\bf{z}} = 1,{\kern 1pt} {\bf{z}} \ge 0} \right\}.
	\end{equation}
	The augmented Lagrangian function for (\ref{P9}) is
	\begin{equation}
		{\cal L}\left( {{\kern 1pt} {\bf{u}},{\bf{z}},{\bm{\eta }}} \right) = {{\bf{u}}^T}{\bf{Vu}} + {\mathbb I}{_{\mathcal C}}\left( {\bf{z}} \right) + {{\bm{\eta }}^T}\left( {{\bf{u}} - {\bf{z}}} \right) + \frac{\rho }{2}{\left\| {{\bf{u}} - {\bf{z}}} \right\|^2},
	\end{equation}
	where ${\rho}$ is a penalty parameter associated with the equality constraint, and ${\bm{\eta}}$ denotes the corresponding dual variable vector. 
	
	The optimization is carried out by iteratively minimizing the augmented Lagrangian function in the following order:
	\begin{subequations}
		\begin{align}
			{{\bf{u}}^{t + 1}} &= \arg {\kern 1pt} \mathop {\min }\limits_{\bf{u}}  {\kern 1pt} {\cal L}\left( {{\kern 1pt} {\bf{u}},{{\bf{z}}^t},{{\bm{\eta }}^t}} \right), \vspace{1ex}\\
			{{\bf{z}}^{t + 1}} &= \arg {\kern 1pt}\mathop {\min }\limits_{\bf{z}} {\kern 1pt} {\cal L}\left( {{\kern 1pt} {{\bf{u}}^{t + 1}},{\bf{z}},{{\bm{\eta }}^t}} \right), \label{zt+1} \vspace{1ex}\\
			{{\bm{\eta }}^{t + 1}} &= {{\bm{\eta }}^t} + \rho \left( {{{\bf{u}}^{t + 1}} - {{\bf{z}}^{t + 1}}} \right),
			\label{eta}
		\end{align}
	\end{subequations}
	where $t$ represents the iteration index. In the following, we describe the specific update process for each variable in the ADMM algorithm.

	\subsubsection{Update ${\bf{u}}$}
	The vector ${{\bf{u}}^{t + 1}}$ is obtained by taking the first-order derivative of the augmented Lagrangian function:
	\begin{equation}
		\frac{{\partial {\cal L}}}{{\partial {\bf{u}}}} = \left( {2{\bf{V}} + \rho {\bf{I}}} \right){\bf{u}} - \rho {{\bf{z}}^t} + {{\bm{\eta }}^t} = {\bf{0}},
	\end{equation}
	which leads to the following closed-form solution:
	\begin{equation}\label{ut+1}
		{{\bf{u}}^{t + 1}} = {\left( {2{\bf{V}} + \rho {\bf{I}}} \right)^{ - 1}}\left( {\rho {{\bf{z}}^t} - {{\bm{\eta }}^t}} \right).
	\end{equation}
	
	\subsubsection{Update ${\bf{z}}$}
	Based on (\ref{zt+1}), the update for ${\bf{z}}$ is obtained by minimizing the augmented Lagrangian with respect to ${\bf{z}}$:
	\begin{equation}
		\begin{aligned}
			\mathbf{z}^{t+1} &= \arg\min_{{{\bf{z}} \in \mathcal{C}} } \, {\mathbb{I}}_{\mathcal{C}}({\bf{z}}) + \left({\bm{\eta}}^t\right)^T {\bf{z}} + \frac{\rho}{2} \left\| {\bf{u}}^{t+1} - {\bf{z}} \right\|^2 \\
			&= \arg\min_{{\bf{z}} \in \mathcal{C}} \, \frac{\rho}{2} \left\| {\bf{z}} - \left( {\bf{u}}^{t+1} + \frac{1}{\rho} {\bm{\eta}}^t \right) \right\|^2 \\
			&= \arg\min_{{\bf{z}} \in \mathcal{C}} \, \frac{\rho}{2} \left\| {\bf{z}} - {\bf{q}}^{t} \right\|^2,
		\end{aligned}
	\end{equation}
	where we omit constant terms irrelevant to $\mathbf{z}$ and we define the intermediate point as
	\begin{equation}\label{qt}
		{\bf{q}}^t = {\bf{u}}^{t+1} + \frac{1}{\rho} {\bm{\eta}}^t.
	\end{equation}
	According to the definition in (\ref{C}), the feasible set $\mathcal{C}$ corresponds to the unit probability simplex\cite{Fast projection}, hence, the update of $\mathbf{z}$ is equivalent to the Euclidean projection of $\mathbf{q}^t$ onto the probability simplex:
	\begin{equation}
		\mathbf{z}^{t+1} = \Pi_{\mathcal{C}}(\mathbf{q}^t),
	\end{equation}
	where $\Pi_{\mathcal{C}}(\cdot)$ denotes the projection operator. This projection problem can be further formulated as
	\begin{equation}\label{projection}
		\min_{\mathbf{z}} \ \frac{1}{2} \left\| \mathbf{z} - \mathbf{q}^t \right\|^2 \quad s.t. \quad \mathbf{1}^T \mathbf{z} = 1,\quad \mathbf{z} \ge 0,
	\end{equation}
	which also is a convex QP problem with linear equality and inequality constraints. The optimal solution can be derived by using the KKT conditions. First, the Lagrangian function of (\ref{projection}) is expressed as
	\begin{equation}
		{\cal L}\left( {{\kern 1pt} {\bf{z}},\nu ,{\bm{\xi }}} \right) = \frac{1}{2}{\left\| {{\bf{z}} - {\bf{q}}} \right\|^2} + \nu \left( {{{\bf{1}}^T}{\bf{z}} - 1} \right) - {{\bm{\xi }}^T}{\bf{z}},
	\end{equation}
	where $\nu$ and ${\bm{\xi }} \ge 0$ are the dual variables. For each $i \in \left\{1,2,\dots,2NK\right\}$, the KKT conditions are given by
	\begin{subequations}
		\begin{align}
			\frac{{\partial {\cal L}}}{{\partial {z_i}}} = {z_i} - {q_i} + \nu  - {\xi _i} &= 0, \label{zi}\vspace{1ex}\\
			\sum\limits_{i = 1}^{2NK} {{z_i}}  = 1,{z_i} &\ge 0,  \vspace{1ex}\\
			{\xi _i} \cdot {z_i} &= 0.
		\end{align}
	\end{subequations}
	Based on (\ref{zi}), ${{z_i}}$ is given by
	\begin{equation}\label{ziii}
		{z_i} = {q_i} - \nu  + {\xi _i},
	\end{equation}
	and substituting it into $\xi_i \cdot z_i = 0$, we further have
	\begin{itemize}
		\item If $z_i > 0$, then $\xi_i = 0$, and thus $z_i = q_i - \nu$.
		\item If $z_i = 0$, then $q_i - \nu + \xi_i = 0$, implying $\xi_i = \nu - q_i \ge 0 $, and hence $ q_i \le \nu$.
	\end{itemize}
	Therefore, the optimal $z_i$ can be compactly expressed as
	\begin{equation}\label{zii}
		z_i = \max(q_i - \nu,\, 0).
	\end{equation}
	
	The equality constraint $\sum_i z_i = 1$ must also be satisfied, i.e.,
	\begin{equation}\label{sumzi}
		\sum\limits_{i = 1}^{2NK} {\max(q_i - \nu,\, 0)}  = 1.
	\end{equation}
	To satisfy this constraint, a unique threshold value $\nu = \theta$ can be chosen such that the identity holds. Since the function $\sum \max(q_i - \nu, 0)$ is monotonically non-increasing with respect to $\nu$, the ``\textit{Linear Time Projection Algorithm}'' proposed in \cite{Efficient projection} can be used to efficiently compute the threshold $\theta$, as outlined here:
	\begin{itemize}[labelwidth=4em, leftmargin=5em, align=parleft]
		\item[\textbf{Step 1:}] Sort the elements of $\mathbf{q}^t$ in descending order: $q_{[1]} \ge q_{[2]} \ge \cdots \ge q_{[2KN]}$.
		\item[\textbf{Step 2:}] Find the largest index $L$ that satisfies
		\begin{equation}\label{step2}
			q_{[L]} - \frac{1}{L} \left( \sum_{j=1}^L q_{[j]} - 1 \right) > 0.
		\end{equation}
		\item[\textbf{Step 3:}] Set the threshold as:
		\begin{equation}
			\theta = \frac{1}{L} \left( \sum_{j=1}^L q_{[j]} - 1 \right).\label{theta}
		\end{equation}
	\end{itemize}
	With $\theta$ obtained as above, each $z_i$ is computed by substituting (\ref{theta}) into (\ref{zii}). This algorithm has a complexity of only $\mathcal{O}(2KN)$ and guarantees that the updated $\mathbf{z}^{t+1}$ lies on the simplex. For more details, please refer to \cite{Efficient projection}.
	
	\subsubsection{Update ${\bm{\eta}}$}
	The vector ${\bm{\eta}}^{t + 1}$ is obtained via (\ref{eta}).
	
	The above steps constitute a simple and efficient ADMM framework, in which each subproblem admits a closed-form or easily computable solution, enabling low complexity iterations. Based on these updates, the complete ADMM algorithm for solving problem $\mathcal{P}_8$ is summarized in \textbf{Algorithm~\ref{Algorithm1}}.
	
	Building on this, three methods can be used to solve problem \(\mathcal{P}_6\):
	\begin{enumerate}
		\item \textbf{CVX-based}: Directly solve the SOCP \(\mathcal{P}_6\) via CVX with a complexity of approximately \(\mathcal{O}((2N_t N)^3)\);
		\item \textbf{QP-based}: Solve the QP \(\mathcal{P}_8\) with \(2KN\) variables using interior-point methods, with complexity of approximately \(\mathcal{O}((2KN)^3)\);
		\item \textbf{ADMM-based}: Solve \(\mathcal{P}_8\) iteratively via \textbf{Algorithm 1}, where the complexity is mainly due to the matrix inversion involved in updating \(\mathbf{u}^{t+1}\), leading to an overall complexity of \(\mathcal{O}(t_{\max}(2KN)^2)\).
	\end{enumerate}
	Note that the the linear time projection used in the ADMM algorithm reduces the per-iteration complexity, while the number of iterations remains acceptable, yielding the lowest overall computational cost and making it particularly suitable for large-scale problems.
	
	\begin{algorithm}[!t]
		\caption{Proposed ADMM algorithm for solving $\mathcal{P}_8$}
		\label{Algorithm1}
		\begin{algorithmic}[1]
			\State \textbf{Input:} $\mathbf{V}$, $\rho$, $t_{\max}$, $\epsilon$.\vspace{0.5ex}
			\State \textbf{Initialization:} Initialize $\mathbf{u}^0$, $\mathbf{z}^0$, $\bm{\eta}^0$, set iteration $t=0$.\vspace{0.5ex}
			\While {$t < t_{\max}$ \textbf{and} $\delta > \epsilon$}\vspace{0.5ex}
			\State Update $\mathbf{u}^{t+1}$ via (\ref{ut+1}).\vspace{0.5ex}
			\State Compute $\mathbf{q}^t$ via (\ref{qt}).\vspace{0.5ex}
			\State \parbox[t]{0.92\linewidth}{
				Obtain $\mathbf{z}^{t+1}$ by projecting $\mathbf{q}^t$ onto the simplex $\mathcal{C}$ using the \textit{Linear Time Projection Algorithm} in (\ref{step2}), (\ref{theta}) and (\ref{zii}).
			} \vspace{0.5ex}
			\State Update $\bm{\eta}^{t+1}$ via (\ref{eta}).\vspace{0.5ex}
			\State $\delta = {\left\| {{\bf{u}}^t - {\bf{z}}^t} \right\|^2}$.\vspace{0.5ex}
			\State $t = t + 1$
			\EndWhile
			\State \textbf{Output:} $\mathbf{u}^\star = \mathbf{u}^t$.
		\end{algorithmic}
	\end{algorithm}
	
	\section{Proposed CI Waveform Design for QAM Modulation} \label{POCIWDQAM}
	In Section \ref{POCIWDPSK}, we have investigated optimal CI-based waveform design for PSK modulated symbols and derived the structure of the optimal closed-form solution. Unlike PSK modulations, where all constellation points can benefit from CI, QAM constellations include some inner points that inherently cannot exploit CI due to their locations within the decision region. As a result, the CI constraints adopted in ${\mathcal{P}}_5$ are not applicable to QAM modulated symbols. To address this issue, below we propose an optimal CI waveform design for QAM constellations and derive the corresponding closed-form solution structure.

	\subsection{Problem Formulation}
\begin{figure}[!t]
		\centering
		\includegraphics[width=1.6 in]{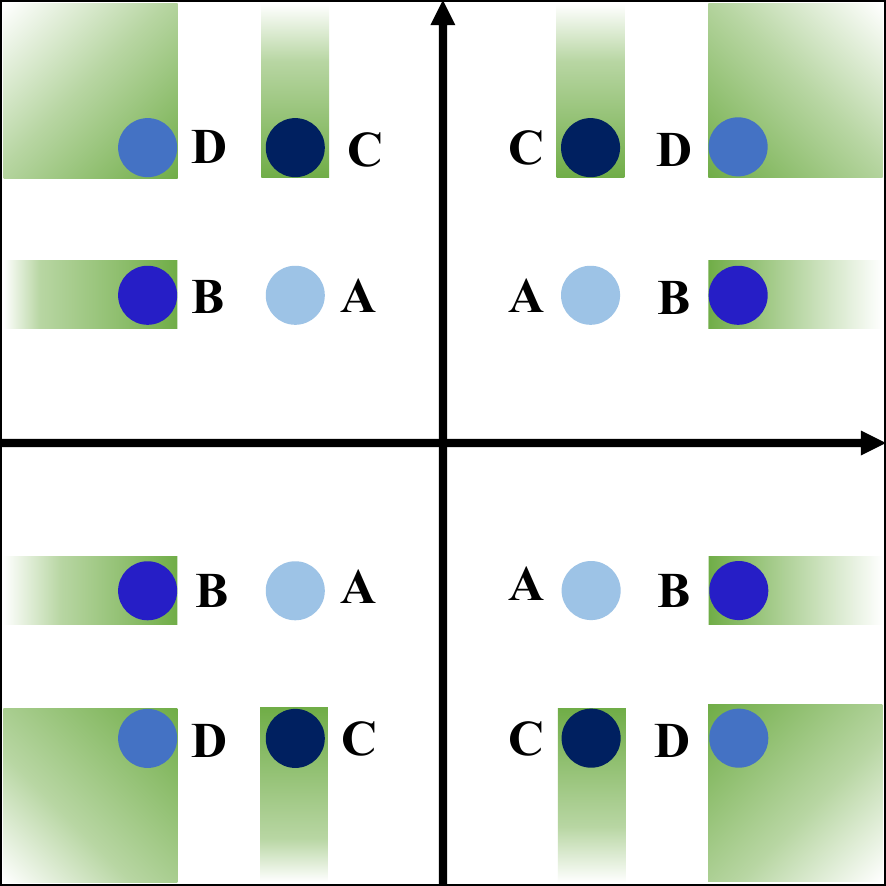}
		\caption{Constellation point categorization and CI metric for 16QAM.}
		\label{FIg3}
	\end{figure}

	As illustrated in Fig. \ref{FIg3}, the inner points ``A'' are located at the center of the 16QAM constellation, where CI cannot be exploited in either the real or imaginary dimensions. In contrast, boundary points ``B'' and ``C'' allow CI exploitation along only one dimension—either the real or the imaginary axis, while points ``D'', located at the outermost corner, can leverage CI in both dimensions. Based on this observation, we reformulate the CI constraints as follows\cite{BLP CI,SLPMLD}:
	\begin{equation}\label{QAMlamda}
		\gamma _{k,n}^\mathfrak{U} \ge t,\gamma _{k,n}^\mathfrak{V} = t,\forall \gamma _{k,n}^\mathfrak{U} \in {\mathfrak{U}},\forall \gamma _{k,n}^\mathfrak{V} \in {\mathfrak{V}},
	\end{equation}
	where the sets $\mathfrak{U}$ and $\mathfrak{V}$ respectively represent the outer points where CI can be exploited, and the inner points where it cannot. The parameters $\gamma _{k,n}^\mathfrak{U}$ and $\gamma _{k,n}^\mathfrak{V}$ are real-valued scalings introduced to denote the CI effect. With (\ref{QAMlamda}), the received noise-free signal for the \textit{k}-th user in the \textit{n}-th symbol slot can be expressed as
	\begin{equation}\label{RS}
		{\bf{h}}_k^T{{\bf{x}}^n} = \gamma _{k,n}^{\frak R} \cdot {\mathcal{R} \left\{ {{s_{k}^n}} \right\}} + \gamma _{k,n}^{\frak J} \cdot {j \cdot {\mathcal J}\left\{ {{s_{k}^n}} \right\}},\forall k \in {\cal K},\forall n \in {\cal N},
	\end{equation}
	where $\gamma _{k,n}^{\frak R}$ and $ \gamma _{k,n}^{\frak J}$ represent the scaling factors for the real and imaginary components, and (\ref{RS}) can be further expressed in a more compact form as
	\begin{equation}
		{\bf{h}}_k^T{{\bf{x}}^n} = {\bm{ \gamma }}_{k,n}^T{{{\bf{\tilde s}}}_{k,n}},\forall k \in {\cal K},\forall n \in {\cal N},
	\end{equation}
	where 
	\begin{equation}\label{gamma&s}
		\begin{array}{l}
			{{\bm{ \gamma }}_{k,n}} = {\left[ {\gamma _{k,n}^{\frak R},\gamma _{k,n}^{\frak J}} \right]^T}$, ${{{\bf{\tilde s}}}_{k,n}} = {\left[ {\mathcal{R} \left\{ {{s_{k}^n}} \right\}},{j \cdot {\mathcal J}\left\{ {{s_{k}^n}} \right\}} \right]^T}.
		\end{array}
	\end{equation}
	
	Following a similar formulation as for ${\mathcal{P}}_5$, the CI-based waveform design problem can be formulated as
	\begin{equation}\label{P10}
		\begin{array}{l}
			{\mathcal{P}}_{10}:\mathop {\max }\limits_{{{\bf{x}}^n}} {\kern 1pt} {\kern 1pt} {\kern 1pt} {\kern 1pt} t \vspace{1ex}\\
			{\kern 27pt} s.t.{\kern 6pt}{\bf{h}}_k^T{{\bf{x}}^n} = {\bm{ \gamma }}_{k,n}^T{{{\bf{\tilde s}}}_{k,n}},\forall k \in {\cal K},\forall n \in {\cal N} \vspace{1ex}\\
			{\kern 48pt} t - \gamma _{m}^{\frak U} \le 0,\forall \gamma _{m}^{\frak U} \in {\frak U} \vspace{1ex} \\
			{\kern 48pt}t - \gamma _{n}^{\frak V} = 0,\forall \gamma _{n}^{\frak V} \in {\frak V} \vspace{1ex} \\
			{\kern 46pt}\sum\limits_{n = 1}^N {\left\| {{{\bf{x}}^n}} \right\|_2^2}  \le N{p_0},
		\end{array}
	\end{equation}
	where $\gamma _{m}^{\frak U}$ and $\gamma _{m}^{\frak V}$ respectively denote the scaling factors that can and cannot exploit CI. We thus have 
	\begin{equation}
		\begin{array}{l}
			{\frak U} \cup {\frak V} = \left\{ {\gamma _{1,1}^{\frak R},\gamma _{1,1}^{\frak J}, \cdots ,\gamma _{k,n}^{\frak R},\gamma _{k,n}^{\frak J}, \cdots ,\gamma _{K,N}^{\frak R},\gamma _{K,N}^{\frak J}} \right\},  \vspace{1ex} \\
			card\left\{ {\frak O} \right\} + card\left\{ {\frak I} \right\} = 2KN.
		\end{array}
	\end{equation}
	${\mathcal{P}}_{10}$ is a convex problem and can be solved by using standard convex solvers in which  the optimization objective $t$ is equal to the minimum value of $\gamma _{k,n}^{\frak R}$ and $\gamma _{k,n}^{\frak J}$.
	
	\subsection{Optimal Waveform $\bf X$ for QAM Modulation}
	
	\textit{Proposition 2}: The optimal waveform ${{\bf{X}}}$ for ${\mathcal{P}}_{10}$ can be expressed as
	\begin{equation}
		{\bf{X}} = {\bf{H}}^T{\left( {{\bf{H}}{{\bf{H}}^T}} \right)^{ - 1}}{{\bf{\bar U}}}{\bf{\Gamma \tilde S}},
	\end{equation}
	where ${{\bf{\bar U}}}$, ${\bf {\Gamma}}$ and ${\bf{ \tilde S}}$ are formulated in Appendix B as (\ref{U}), (\ref{Gamma}) and (\ref{sbar}), respectively.
	
	\textit{Proof}: See Appendix B.
	
	Similar to the PSK modulation scenario, the block-level power constraint is active when optimality is achieved, which indicates that
	\begin{equation}\label{gammaa}
		\begin{aligned}
			\sum\limits_{n = 1}^N {\left\| {{{\bf{x}}^n}} \right\|_2^2} &= N{p_0}\\
			&\Rightarrow \sum\limits_{n = 1}^N {\left\| {{\bf{G\bar U}}{\rm{diag}}\left( {{{\bm{\gamma }}^n}} \right){{{\bf{\tilde s}}}^n}} \right\|_2^2}  = N{p_0}\\
			&\Rightarrow \sum\limits_{n = 1}^N {\left( {{\bm{\gamma }}_n^T{\bf{\tilde G}}_n^H{{{\bf{\tilde G}}}_n}{{\bm{\gamma }}_n}} \right)}  = N{p_0}\\
			&\Rightarrow {{\bm{\gamma }}^T}{{{\bf{\tilde G}}}^H}{\bm{\tilde G\gamma }} = N{p_0},
		\end{aligned}
	\end{equation}
	where ${\bf{\tilde G}}_n = {\bf{G\bar U}}{\rm{diag}}\left( {{{{\bf{\bar s}}}^n}} \right) \in \mathbb{C} {^{N_T \times 2K}}$, ${\bf{\tilde s}}^n = {\left[ {{\bf{\tilde s}}_{1,n}^T ,{\bf{\tilde s}}_{2,n}^T, \cdots ,{\bf{\tilde s}}_{K,n}^T} \right]^T} \in \mathbb{C} {^{2K \times 1}}$, ${\bm{\gamma }} = {\left[ { ({{\bm{\gamma }}^1})^T,({{\bm{\gamma }}^2})^T, \cdots ,({{\bm{\gamma }}^N})^T} \right]^T} \in \mathbb{R} {^{2KN \times 1}}$ and $\tilde{\mathbf{G}} = \operatorname{blkdiag}(\tilde{\mathbf{G}}_1, \tilde{\mathbf{G}}_2, \ldots, \tilde{\mathbf{G}}_N) \in \mathbb{C}^{NN_T \times 2NK}$. Substituting (\ref{gammaa}) in ${\mathcal{P}}_{10}$ leads to the following reformulation of the problem:
	\begin{equation}\label{P11}
		\begin{array}{l}
			{\mathcal{P}}_{11}:\mathop {\max }\limits_{\bm{\gamma }} {\kern 1pt} {\kern 1pt} {\kern 1pt} {\kern 1pt} t \vspace{1ex}\\
			{\kern 27pt} s.t.{\kern 6pt}{{\bm{\gamma }}^T}{{{\bf{\tilde G}}}^H}{\bm{\tilde G\gamma }} - N{p_0} = 0 \vspace{1ex}\\
			{\kern 48pt} t - \gamma _{m}^{\frak U} \le 0,\forall \gamma _{m}^{\frak U} \in {\frak U} \vspace{1ex} \\
			{\kern 48pt}t - \gamma _{n}^{\frak V} = 0,\forall \gamma _{n}^{\frak V} \in {\frak V}.
		\end{array}
	\end{equation}
	
	\textit{Proposition 3}: The optimal value of ${\bm{\gamma }}$ is found by solving the following QP problem and substituting the solution for ${\bm{\varphi }}$ into (\ref{GAMMMA}) and (\ref{dsad}):
	\begin{equation}\label{P12}
		\begin{array}{l}
			{\mathcal{P}}_{12}:\mathop {\min }\limits_{\bm{\varphi }} {\kern 3pt} {{\bm{\varphi }}^T}{{{\bf{\tilde V}}}^{ - 1}}{\bm{\varphi }} \vspace{1ex}\\
			{\kern 27pt}s.t.{\kern 3pt}{{\bf{1}}^T}{\bm{\varphi }} = 1 \vspace{1ex}\\
			{\kern 45pt}{{\hat \varphi }_i} \ge 0,\forall i \in \left\{ {1,2, \cdots ,2KN} \right\}.
		\end{array}
	\end{equation}
	
	\textit{Proof}: See Appendix C.
	
	The above derivation reveals that the QAM modulation problem shares a similar structure with the PSK case, and the optimal transmit waveform also admits a closed-form solution. Furthermore, ${\mathcal{P}}_{12}$ has the same form as ${\mathcal{P}}_{8}$, except that $\mathbf{V}$ is replaced by ${{\mathbf{\tilde V}}^{-1}}$, and both $\mathbf{V}$ and ${{\mathbf{\tilde V}}^{-1}}$ are symmetric real matrices. This implies that ${\mathcal{P}}_{12}$ can also be efficiently solved by using the proposed \textbf{Algorithm 1} in the QAM case, resulting in a significant reduction in computational complexity.
	
	
	\section{Simulation Results}\label{Simulation}	
	This section presents simulation results to verify the effectiveness of the proposed waveform design approach and compares its performance with several conventional methods to demonstrate its advantages. In the simulation setup, each element of the channel matrix $\bf H$ follows a standard complex Gaussian distribution: ${{\bf H}_{m,n}} \in \mathcal{CN}(0,1)$. The transmit power is set to be $p_0 = 1 (\text{W})$ for each symbol slot, resulting in a total block power budget $P_{total} = N \cdot p_0 = N (\text{W}) $. The proposed nonlinear waveform optimization algorithm is compared against traditional methods including ZF, RZF, CI-SLP, and CI-BLP. The presented results are averaged over 10000 independent channel realizations. All runtime evaluations are carried out on a Windows 11 platform with an Intel i9-12950HX CPU and 64 GB RAM.
	
	The abbreviations of the algorithms involved in the simulation figures of this section are summarized as follows:
	\begin{enumerate}
		\item ``ZF'': Traditional ZF precoding with block-level power budget \cite{ZF};
		\item ``RZF'': Traditional RZF precoding with block-level power budget \cite{RZF};
		\item ``CI-SLP'': Symbol-level CI precoding with symbol-level power constraint for PSK \cite{CF-PSK} and QAM \cite{CF-QAM};
		\item ``CI-BLP'': Block-level CI precoding with block-level power constraint for PSK and QAM \cite{BLP CI};
		\item ``Nonlinear-X-CVX'': CVX-based solution to original waveform problems ${\mathcal{P}}_{6}$ and ${\mathcal{P}}_{10}$;
		\item ``Nonlinear-X-QP'': Proposed QP-based solutions to ${\mathcal{P}}_{8}$ and ${\mathcal{P}}_{12}$ ;
		\item ``Nonlinear-X-ADMM'': Proposed ADMM-based \textbf{Algorithm 1} for solving ${\mathcal{P}}_{8}$ and ${\mathcal{P}}_{12}$.
	\end{enumerate}

\begin{figure}[!t]
		\centering
		\includegraphics[width=2.5in]{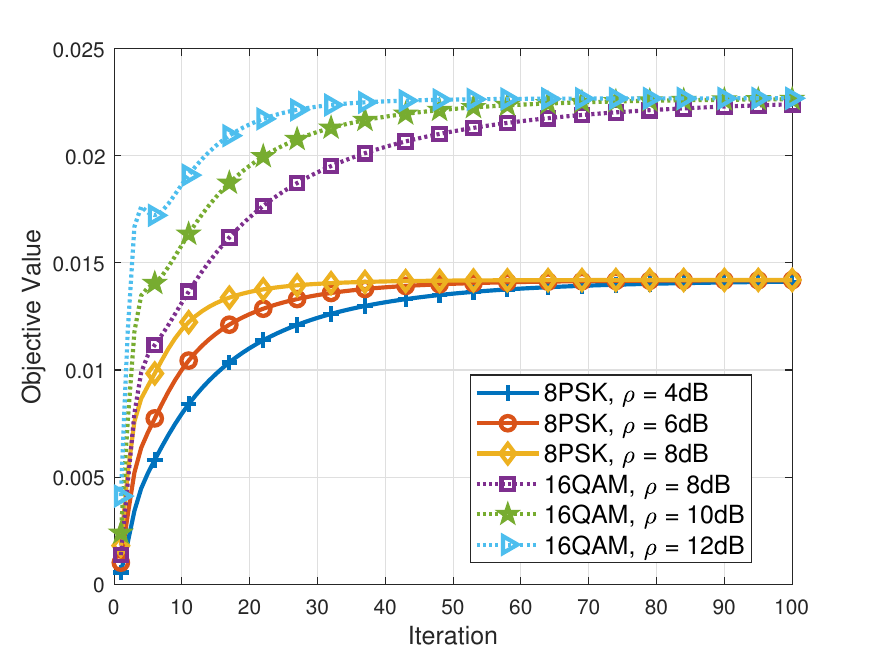}
		\caption{Convergence of the proposed ADMM algorithm for PSK and QAM modulations with different penalty parameters $\rho$.}
		\label{convergence}
	\end{figure}

	Fig. \ref{convergence} shows the convergence of the proposed ADMM algorithm for PSK and QAM modulations with different penalty parameters \(\rho\). The objective functions are \( \mathbf{u}^T \mathbf{V} \mathbf{u} \) for PSK and \( \mathbf{u}^T \mathbf{\tilde{V}}^{-1} \mathbf{u} \) for QAM.  It can be observed that with moderate \(\rho\), the proposed ADMM algorithm converges within approximately 20 to 30 iterations. However, excessively large \(\rho\) values may cause oscillations and reduce solution accuracy. For 16QAM, larger \(\rho\) is typically needed due to its higher constellation complexity.
	
\begin{figure}[!t]
		\centering
		\includegraphics[width=2.5in]{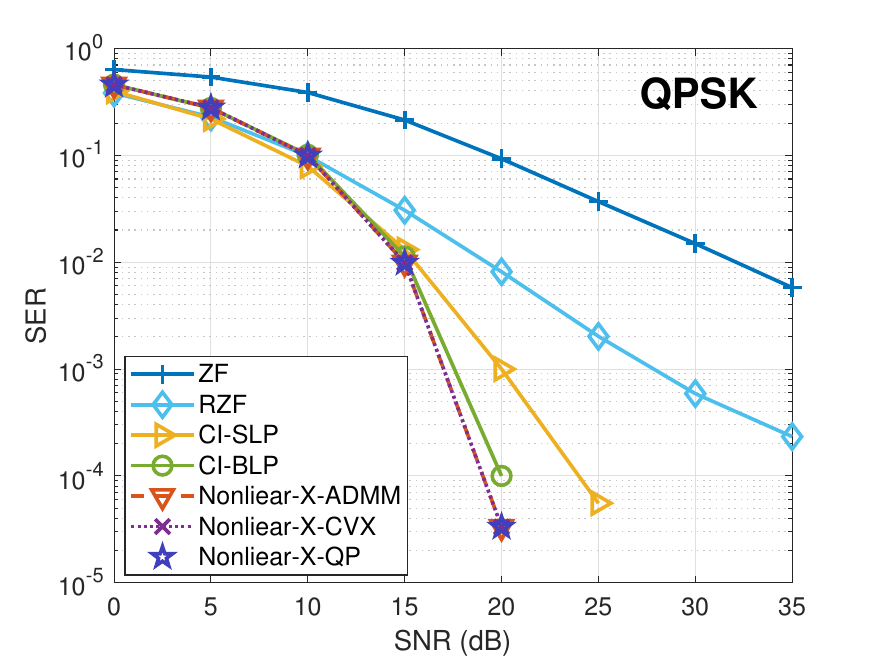}
		\caption{SER v.s. SNR, QPSK, ${N_T}=12, K=12$ and $N=15$.}
		\label{QPSK1}
	\end{figure}

\begin{figure}[!t]
		\centering 
		\includegraphics[width=2.5in]{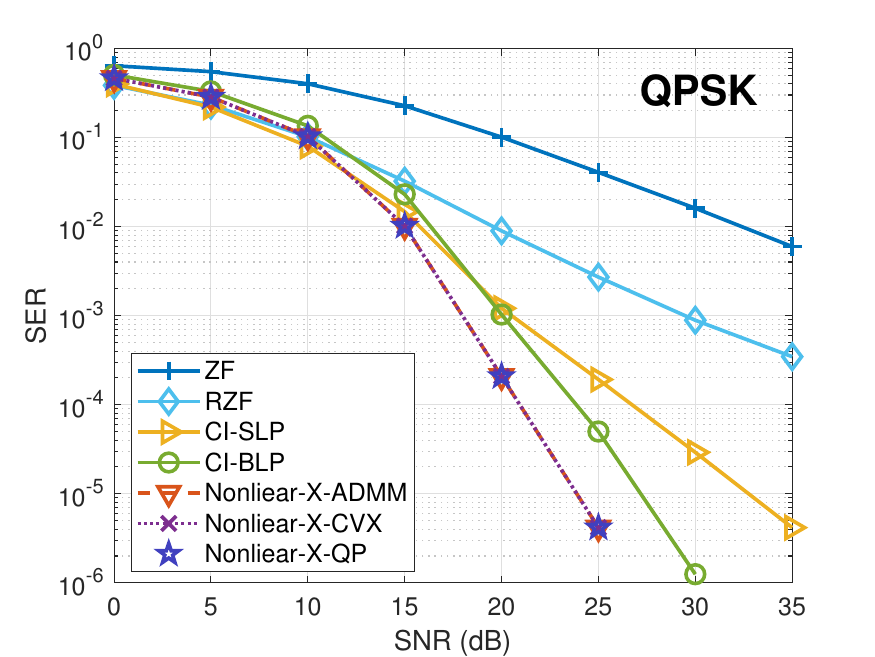}
		\caption{SER v.s. SNR, QPSK, ${N_T}=12, K=12$ and $N=40$.}
		\label{QPSK2}
	\end{figure}

	Figs. \ref{QPSK1} and \ref{QPSK2} show the SER performance versus SNR for QPSK modulation, with 12 transmit antennas and 12 users. The block length is set to either $N = 15$ or $N = 40$. In both cases, all CI-based schemes significantly outperform conventional ZF and RZF approaches. Among them, the proposed ``Nonlinear-X'' waveform design achieves the best SER performance, clearly demonstrating its advantage. Comparing these two figures, CI-BLP shows degraded performance as the block length increases. This is because CI-BLP suffers from a reduced number of DoFs. Notably, when $N = 40$, the proposed ``Nonlinear-X'' schemes achieve even greater SER gains, highlighting their robustness. The three implementations: ``Nonlinear-X-CVX'', ``Nonlinear-X-QP'' and ``Nonlinear-X-ADMM'' yield identical SER results, validating the correctness of the closed-form structure and the efficiency of the equivalent QP formulation and the ADMM-based algorithm.
	
\begin{figure}[!t]
		\centering 
		\includegraphics[width=2.5in]{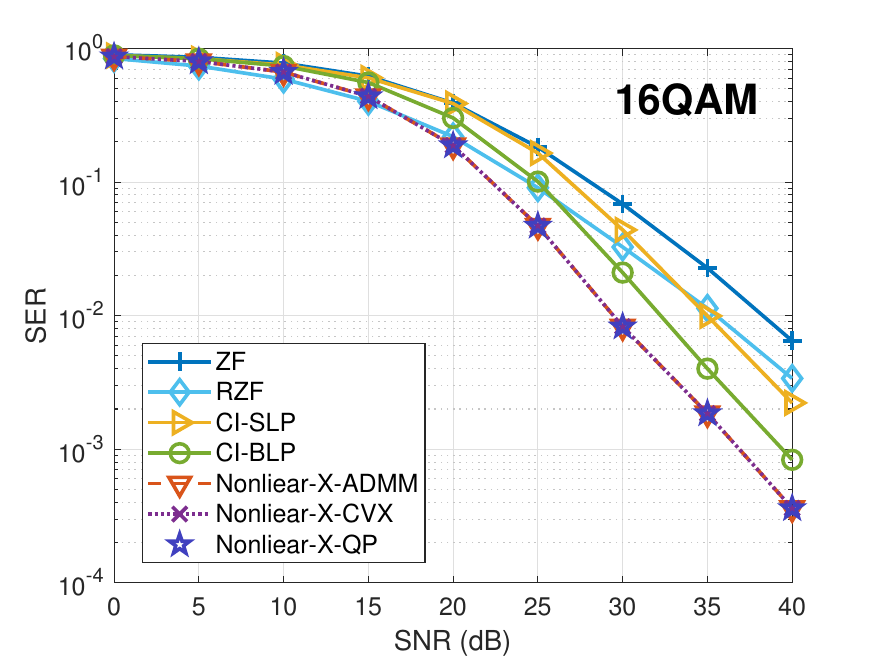}
		\caption{SER v.s. SNR, 16QAM, ${N_T}=12, K=12$ and $N=15$.}
		\label{16QAM1}
	\end{figure}

\begin{figure}[!t]
		\centering 
		\includegraphics[width=2.5in]{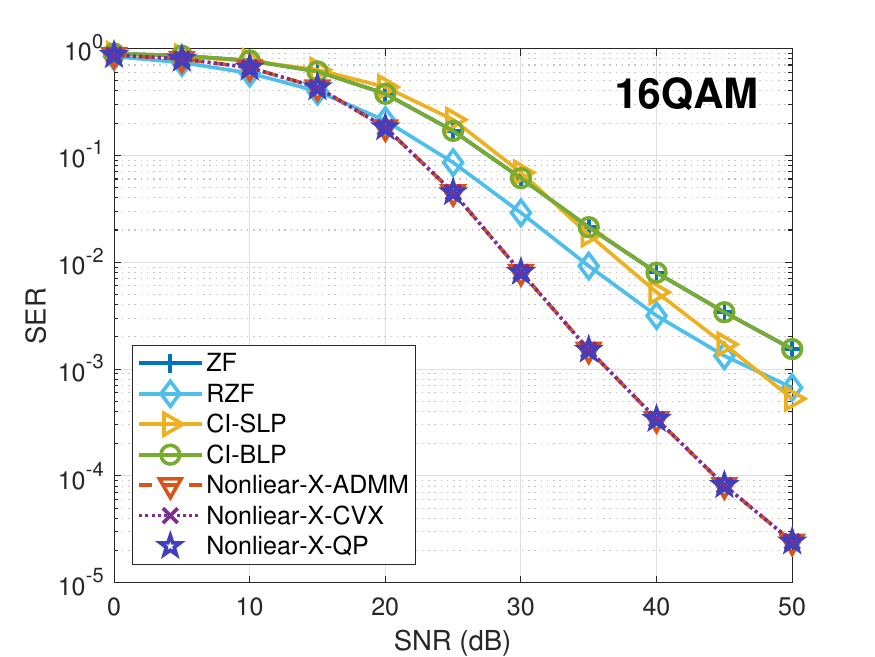}
		\caption{SER v.s. SNR, 16QAM, ${N_T}=12, K=12$ and $N=40$.}
		\label{16QAM2}
	\end{figure}

	Figs. \ref{16QAM1} and \ref{16QAM2} show the SER performance for 16QAM modulation with $N_T = K = 12$ and block lengths $N = 15$ and $N = 40$, respectively. When $N = 15$, all CI-based approaches outperform ZF and RZF, but for $N=40$, the performance of CI-BLP degrades to that of ZF due to its limited design flexibility. Similarly, CI-SLP shows only marginal gains over ZF and even performs worse than RZF, revealing its limitations under higher-order modulation and large block lengths. In contrast, the proposed ``Nonlinear-X'' approaches maintain strong performance for all settings and achieves significant SER gains at $N = 40$, highlighting its robustness and superior CI exploitation capability.
	
\begin{figure}[!t]
		\centering 
		\includegraphics[width=2.5in]{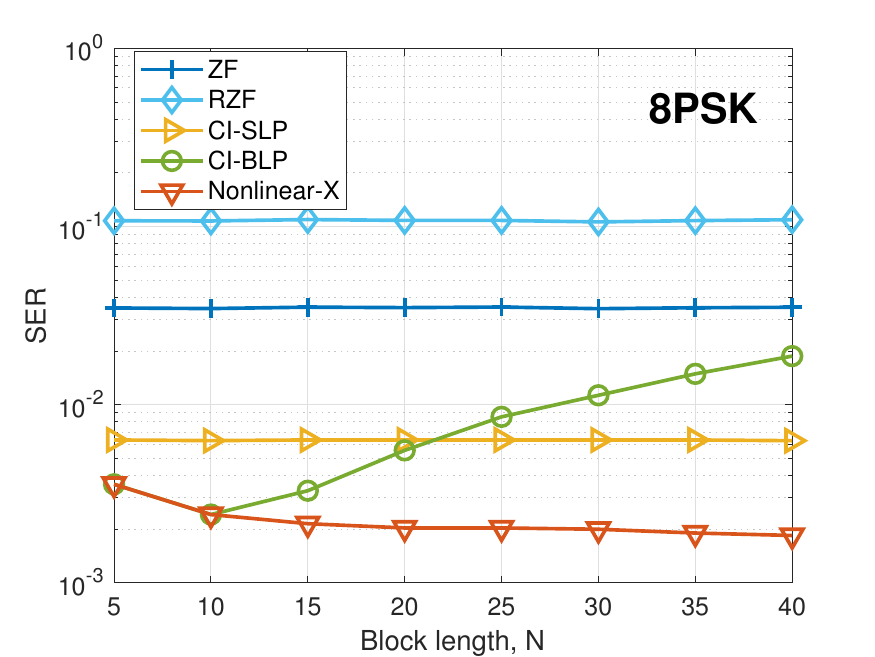}
		\caption{SER v.s. block length $N$, 8PSK, ${N_T}=K=12$ and $\text{SNR}=25\text{dB}$.}
		\label{SERVSN}
	\end{figure}

	Fig. \ref{SERVSN} shows the impact of block length on SER performance for 8PSK modulation with a transmit SNR of $25\text{dB}$. The SER of ZF, RZF, and CI-SLP remains constant across block lengths, as these methods are independent of the block lengths. With ${N_T}=K=12$, CI-BLP and ``Nonlinear-X'' exhibit identical SER when the block length is less than 12 due to sufficient DoFs in CI-BLP approach. As the block length increases, CI-BLP suffers performance degradation, eventually underperforming CI-SLP, due to insufficient DoFs. In contrast, the proposed ``Nonlinear-X'' consistently outperforms all benchmarks. Its SER initially decreases with the increase of block lengths and eventually saturates when the block length becomes sufficiently large, demonstrating the effectiveness and performance superiority of the proposed approach.
		
\begin{figure}[!t]
		\centering 
		\includegraphics[width=2.5in]{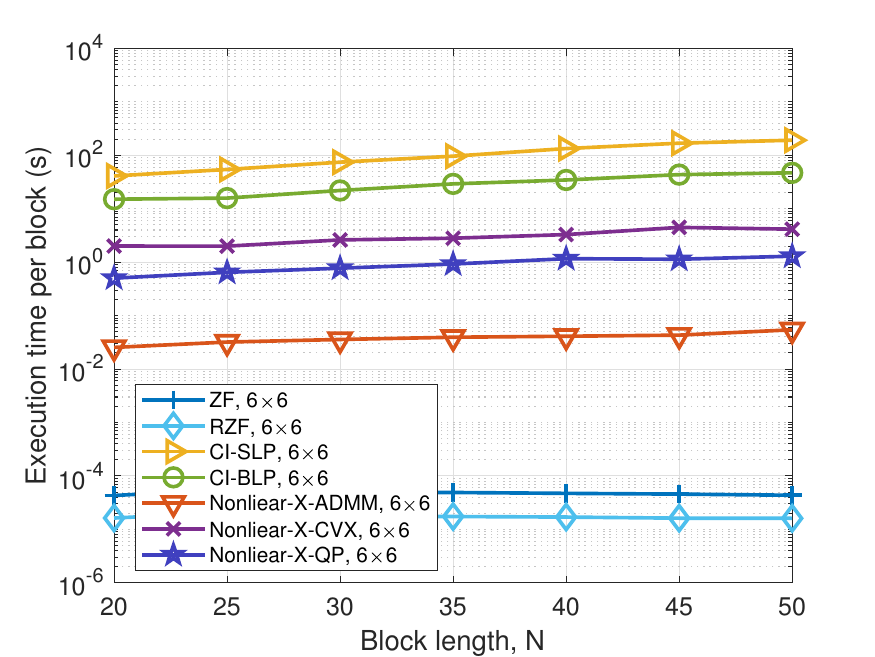}
		\caption{Execution time v.s. block length $N$.}
		\label{Execution time}
	\end{figure}

	Fig. \ref{Execution time} depict the computational complexity of the proposed algorithms in terms of execution time, where the system dimensions are set to $6 \times 6 $. The execution time increases with block length for all CI-based approaches, as more symbols are jointly optimized, leading to higher computational burden. The ``Nonlinear-X-ADMM'' method consistently exhibits the lowest complexity, primarily due to the application of the linear-time projection algorithm within the ADMM framework, which reduces the per-iteration cost while maintaining an acceptable number of iterations.

	\section{Conclusion}\label{Conclusion}
	This paper has proposed an optimal CI-based nonlinear precoding framework for MU-MISO systems under both PSK and QAM modulations. By formulating the corresponding Lagrangian functions and applying the resulting KKT conditions, closed-form solutions were obtained for the proposed problems. Moreover, the original problems were equivalently reformulated as simpler QP problems and an efficient ADMM-based algorithm together with a linear-time projection method were employed to find a low-complexity solution. Compared with conventional CI-SLP and CI-BLP approaches, the proposed method eliminates structural limitations and achieves higher CI gains across the entire block. Numerical results validate its performance advantage, especially for large block lengths and high-order modulations. Future research may focus on extending the CI framework to broader application scenarios, such as ISAC or IRS-assisted systems, to more fully exploit CI gains under diverse system constraints and objectives.
	
	\appendices
	\section{Proof for Proposition 1}
	\noindent \textbf{1. From \(\mathcal{P}_5\) to \(\mathcal{P}_6\):}
	
	Suppose \(\{\mathbf{P}_n^\star\}, t^\star\) are the optimal solutions for \(\mathcal{P}_5\). Defining
	\begin{equation}
		\mathbf{X}^\star = \mathbf{P}^\star \mathbf{\bar{S}} = \big[ \mathbf{P}_1^\star \mathbf{s}^1, \mathbf{P}_2^\star \mathbf{s}^2, \ldots, \mathbf{P}_N^\star \mathbf{s}^N \big],
	\end{equation}
	it follows that
	\begin{equation}
		\mathbf{H} \mathbf{X}^\star = \mathbf{H} \mathbf{P}^\star \mathbf{\bar{S}}.
	\end{equation}
	Since for all \(k,n\),
	\begin{equation}
		\mathbf{h}_k^T \mathbf{P}_n^\star \mathbf{s}^n = \lambda_{k,n} s_k^n,
	\end{equation}
	we have
	\begin{equation}
		\mathbf{H} \mathbf{X}^\star = \mathbf{\Lambda} \mathbf{\bar{S}}.
	\end{equation}
	The power constraint is satisfied for $\mathbf{X}^\star$ since
	\begin{equation}
		\|\mathbf{X}^\star\|_F^2 = \|\mathbf{P}^\star \mathbf{\bar{S}}\|_F^2 \leq N p_0,
	\end{equation}
	and the CI constraints are identical. Therefore, \((\mathbf{X}^\star, t^\star)\) is feasible for \(\mathcal{P}_6\).
	
	\bigskip
	
	\noindent \textbf{2. From \(\mathcal{P}_6\) to \(\mathcal{P}_5\):}
	
	Suppose \(\mathbf{X}^\star, t^\star\) are the optimal solutions for \(\mathcal{P}_6\). We construct \(\mathbf{P}_n^\star\) for \(n=1,\ldots,N\) as
	\begin{equation}
		\mathbf{P}_n^\star \mathbf{s}^n = \mathbf{x}_n^\star, \label{Pn}
	\end{equation}
	where 
	\begin{equation}
		\mathbf{P}_n^\star = \frac{\mathbf{x}_n^\star \cdot \left( \mathbf{s}^n \right)^H}{\|\mathbf{s}^n\|^2},
	\end{equation}
	which leads to
	\begin{equation}
		\mathbf{P}^\star \mathbf{\bar{S}} = \mathbf{X}^\star.
	\end{equation}
	This ensures that both the CI constraints and the power budget are satisfied. Therefore, \(\mathbf{P}_n^\star\) is feasible for \(\mathcal{P}_5\).
	
	\bigskip
	
	\noindent \textbf{3. Equivalence of Optimal Values:}
	
	Because the feasible sets of \(\mathcal{P}_5\) and \(\mathcal{P}_6\) are in one-to-one correspondence via \(\mathbf{X} = \mathbf{P} \mathbf{\bar{S}}\), and the objective function \(t\) is identical, the optimal values coincide:
	\begin{equation}
		\max_{\mathbf{X}} \mathcal{P}_6 = \max_{\{\mathbf{P}_n\}} \mathcal{P}_5,
	\end{equation}
	and their corresponding optimal solutions satisfy
	\begin{equation}
		\mathbf{X}^\star = \mathbf{P}^\star \mathbf{\bar{S}}.
	\end{equation}
	Therefore, the two problems \(\mathcal{P}_5\) and \(\mathcal{P}_6\) are equivalent in terms of feasible set and optimal value. This completes the proof.
	
	\section{Proof for Proposition 2}
	The Lagrangian function of ${\mathcal{P}}_{10}$ is expressed as
	\begin{equation}\label{LQ1}
		\begin{aligned}
			&{\cal L}\left( {{\kern 1pt} {{\bf{x}}^n},t,{\vartheta _{k,l}},{\varphi _m},{\phi _n},{\xi _0}} \right)  \\
			&=  - t + \sum\limits_{n = 1}^N {\sum\limits_{k = 1}^K {{\vartheta _{k,l}}\left( {{\bf{h}}_k^T{{\bf{x}}^n} - {\bm{\gamma }}_{k,n}^T{{{\bf{\tilde s}}}_{k,n}}} \right)} } \\
			&+ \sum\limits_{m = 1}^{card\left\{ {\frak U} \right\}} {{\varphi _m}\left( {t - \gamma _m^{\frak U}} \right)}  + \sum\limits_{n = 1}^{card\left\{ {\frak V} \right\}} {{\phi _n}\left( {t - \gamma _n^{\frak V}} \right)} \vspace{1ex}\\ &+ {\xi _0}\left( {\sum\limits_{n = 1}^N {\left\| {{{\bf{x}}^n}} \right\|_2^2}  - N{p_0}} \right),
		\end{aligned}
	\end{equation}
	where ${\vartheta _{k,l}}$, ${\varphi _m} \ge 0$, ${\phi _n}$ and ${\xi _0} \ge 0$ are the dual variables. Note that ${\vartheta _{k,l}}$ and ${\phi _n}$ may be complex since they are the dual variables with respect to the equality constraints. Based on (\ref{LQ1}), the KKT conditions are derived as
	\begin{subequations}
		\begin{align}
			\dfrac{{\partial {\cal L}}}{{\partial t}} =  - 1 + \sum\limits_{m = 1}^{card\left\{ {\frak U} \right\}} {{\varphi _m}}  + \sum\limits_{n = 1}^{card\left\{ {\frak V} \right\}} {{\phi _n}}  &= 0, \vspace{1ex}\\ 
			\dfrac{{\partial {\cal L}}}{{\partial {{\bf{x}}^n}}} = \sum\limits_{k = 1}^K {{\vartheta _{k,l}}{{\bf{h}}_k}}  + 2{\xi _0}{{\bf{x}}^n} &= {\bf{0}},\forall n \in {\cal N}, \label{KKT6} \vspace{1ex}\\ 
			t - \gamma _n^{\frak V} &= 0,\forall \gamma _n^{\frak V} \in {\frak V},  \vspace{1ex}\\
			{\xi _0}\left( {\sum\limits_{n = 1}^N {\left\| {{{\bf{x}}^n}} \right\|_2^2}  - N{p_0}} \right) &= 0.
		\end{align}
	\end{subequations}
	It can be observed from (\ref{KKT6}) that ${\xi _0} \ne 0$, which leads to ${\xi _0} > 0$. Therefore, ${{\bf{x}}^n}$ can be derived as
	\begin{equation}
		\begin{array}{l}
			\sum\limits_{k = 1}^K {{\vartheta _{k,n}}{{\bf{h}}_k}}  + 2{\xi _0}{{\bf{x}}^n} = {\bf{0}}\\
			\Rightarrow {{\bf{x}}^n} =  - \dfrac{1}{{2{\xi _0}}}\sum\limits_{k = 1}^K {{\vartheta _{k,n}}{{\bf{h}}_k}} \\
			\Rightarrow {{\bf{x}}^n} = \sum\limits_{k = 1}^K {\zeta _k^n{{\bf{h}}_k}}  = {{\bf{H}}^T}{{\bm{\zeta }}^n},
		\end{array}
	\end{equation}
	where we introduce auxiliary variable $\zeta _k^n =  - \dfrac{{{\vartheta _{k,n}}}}{{2{\xi _0}}}$ and auxiliary vector ${{\bm{\zeta }}^n} = {\left[ {\zeta _1^n,\zeta _2^n, \cdots ,\zeta _K^n} \right]^T} \in \mathbb{C} {^{K \times 1}}$. The transmit waveform can be further expressed as
	\begin{equation}
		\begin{aligned}
			{\bf{X}} &= \left[ {{\bf{x}}^1, {\bf{x}}^2, \cdots, {\bf{x}}^N} \right] \vspace{1ex} \\
			&= \left[ {{\bf{H}}^T {\bm{\zeta}}^1, {\bf{H}}^T {\bm{\zeta}}^2, \cdots, {\bf{H}}^T {\bm{\zeta}}^N} \right] \vspace{1ex} \\
			&= {\bf{H}}^T {\bf{Z}}.
		\end{aligned}
	\end{equation}
	According to the first constraint of ${\mathcal{P}}_{10}$, we have
	\begin{equation}
		\begin{aligned}
			{\bf{HX}} &= {{\bf{\bar U}}}diag\left( {\bf{\Gamma }} \right){\bf{\tilde S}} \vspace{1ex}\\
			&\Rightarrow {\bf{H}}{{\bf{H}}^T}{\bf{Z}} = {{\bf{\bar U}}}{\bf{\Gamma }}{\bf{\tilde S}} \vspace{1ex}\\
			&\Rightarrow {\bf{Z}} = {\left( {{\bf{H}}{{\bf{H}}^T}} \right)^{ - 1}}{{\bf{\bar U}}}{\bf{\Gamma }}{\bf{\tilde S}},
		\end{aligned}
	\end{equation}
	where 
	\begin{subequations}
		\begin{align}
			{{\bf{\bar U}}} &= {{\bf{I}}_{K}} \otimes \left[ {1,1} \right]\in \mathbb{R} {^{K \times 2K}}, \label{U} \vspace{1ex}\\ 
			{\bf{\Gamma }} &= \left[ {{\rm{diag}}\left( {{{\bm{\gamma }}^1}} \right),{\rm{diag}}\left( {{{\bm{\gamma }}^2}} \right), \cdots ,{\rm{diag}}\left( {{{\bm{\gamma }}^N}} \right)} \right] \in {\mathbb{R}^{2K \times 2KN}}, \label{Gamma} \vspace{1ex}\\
			{{\bm{\gamma }}^n} &= \left[ {{\bm{\gamma }}_{1,n}^T,{\bm{\gamma }}_{2,n}^T, \cdots ,{\bm{\gamma }}_{K,n}^T} \right] \in {\mathbb{R}^{2K \times 1}},\vspace{1ex}\\
			\tilde{\mathbf{S}} &= \operatorname{blkdiag}(\tilde{\mathbf{s}}^1, \tilde{\mathbf{s}}^2, \ldots, \tilde{\mathbf{s}}^N) \in \mathbb{C}^{2KN \times N}. \label{sbar}
		\end{align}
	\end{subequations}
	Finally, the closed-form solution for ${\bf{X}}$ can be expressed as
	\begin{equation}
		{\bf{X}} = {\bf{H}}^T{\left( {{\bf{H}}{{\bf{H}}^T}} \right)^{ - 1}}{{\bf{\bar U}}}{\bf{\Gamma}}{\bf{\tilde S}},
	\end{equation}
	which completes the proof.
	
	\section{Proof for Proposition 3}
	The Lagrangian function of ${\mathcal{P}}_{11}$ is formulated as
	\begin{equation}\label{L2}
		\begin{array}{l}
			{\cal L}\left( {{\kern 1pt} {\bm{\gamma }},t,{\rho _0},{{\hat \varphi }_m},{{\hat \phi }_n}} \right) =  - t + {\rho _0}\left( {{{\bm{\gamma }}^T}{\bf{\tilde T }}{\bm \gamma} - N{p_0}} \right)  \vspace{1ex}\\
			\sum\limits_{m = 1}^{card\left\{\frak U \right\}} {{\hat \varphi _m}\left( {t - \gamma _m^{\frak U}} \right)}  + \sum\limits_{n = 1}^{card\left\{\frak V \right\}} {{\hat \phi _n}\left( {t - \gamma _n^{\frak V}} \right)},
		\end{array}
	\end{equation}
	where ${\bf{\tilde T }} ={\cal{R}} \left({{{\bf{\tilde G}}}^H}{\bf{\tilde G }}\right)$. This function has the same form as (78) in \cite{CF-QAM}, therefore, it can be further simplified as
	\begin{equation}
		\begin{array}{l}
			{\cal L}\left( {{\kern 1pt} {\bm{ \gamma }},t,{\rho _0},{{\bm{\varphi}}}} \right) = \left( {{{\bf{1}}^T}{{\bm{\varphi}}} - 1} \right)t - {{\bm{\varphi}}}^T{\bm{ \gamma }} \vspace{1ex}\\
			+ {\rho _0}{{{\bm{\gamma }}}^T}{\bf{E}}{\bf{\tilde T }}{{\bf{E}}^T}{\bm{\gamma }} - {\rho _0}N{p_0},
		\end{array}
	\end{equation}
	where ${{\bm{\varphi}}}$ is the dual variable vector given by
	\begin{equation}
		{\bf{\varphi }} = {\left[ {{{\hat \varphi }_1}, \cdots {{\hat \varphi }_{card\{\frak U\} }},{{\hat \phi }_1}, \cdots ,{{\hat \phi }_{card\{\frak V\} }}} \right]^T},
	\end{equation}
	${\bf{E}} = { \left[ {\bf{e}}_{L ({{\tilde s}_1})}, {\bf{e}}_{L ({{\tilde s}_2})} , \cdots , {\bf{e}}_{L ({{\tilde s}_{2KN}})} \right]^T}$, and $ L(\cdot) $ is a ``\textbf{Locater}'' function that returns the index of ${{\tilde s}_m}$ in ${\bf{{\bar s}}}$, defined as
	\begin{equation}
		L({\tilde s}_m) = k, {\kern 2pt}\text{if}{\kern 4pt} {\tilde s}_m = {\bar s}_k.
	\end{equation}
	The invertible matrix ${\bf{E}}\in \mathbb{R} {^{2KN \times 2KN}}$ is introduced to rearrange the columns and rows of the matrix ${\bf{\tilde T }}$ for notational and mathematical convenience. 
	
	Subsequently, the KKT conditions for (\ref{L2}) can be expressed as
	\begin{subequations}
		\begin{align}
			\dfrac{{\partial {\cal L}}}{{\partial t}}  = {{{\bf{1}}^T}{\bm{\varphi }} - 1} &= 0, \vspace{1ex}\\ 
			\frac{{\partial {\cal L}}}{{\partial {\bm{ \gamma }}}} = 2{\rho _0}{\bf{\tilde V }}{\bm{\hat \gamma}} - {\bm{\varphi }} &= {\bf{0}}, \label{KKT7} \vspace{1ex}\\ 
			{{{\bm{\gamma }}}^T}{\bf{\tilde V }}{\bm{\gamma }} - N{p_0} &= 0,  \label{KKT8}
		\end{align}
	\end{subequations}
	where ${\bf{\tilde V}} = {\bf{E}}{\bf{\tilde T }}{{\bf{E}}^T}$. Based on (\ref{KKT7}), we obtain
	\begin{equation}\label{GAMMMA}
		{\bm{\gamma }} = \frac{1}{{2{\rho _0}}}{{{\bf{\tilde V}}}^{ - 1}}{\bm{\varphi }},
	\end{equation}
	and substituting (\ref{GAMMMA}) into (\ref{KKT8}), allows us to have
	\begin{equation}\label{dsad}
		{\rho _0} = \sqrt {\frac{{{{\bm{\varphi }}^T}{{{\bf{\tilde V}}}^{ - 1}}{\bm{\varphi }}}}{{4N{p_0}}}}.
	\end{equation}
	It can be observed that ${\mathcal{P}}_{11}$ satisfies Slater's condition, and the dual problem can be further expressed as
	\begin{equation}
		\begin{aligned}
			{\cal \tilde U} &= \mathop {\max }\limits_{{\bm{\varphi }},{\rho _0}} {\kern 1pt} {\kern 1pt} {\kern 1pt} \mathop {\min }\limits_{{\bm{\gamma }},t} {\kern 1pt} {\kern 1pt} {\kern 1pt} {\cal L}\left( {{\kern 1pt} {\bm{\gamma }},t,{\rho _0},{{\bm{\varphi}}}} \right) \vspace{1ex}\\
			&= \mathop {\max }\limits_{{\bm{\varphi }},{\rho _0}} {\kern 1pt} {\kern 1pt} {\kern 1pt} {{\bm{\varphi}}}^T{\bm{ \gamma }}+ {\rho _0}{{{\bm{ \gamma }}}^T}{\bm{\varphi }}{\bm{ \gamma }} - {\rho _0}N{p_0} \vspace{1ex}\\
			&= \mathop {\max }\limits_{{\bm{\varphi }}} {\kern 1pt} {\kern 1pt} {\kern 1pt}  - \sqrt {N{p_0} {{\bm{\varphi }}^T}{{{\bf{\tilde V}}}^{ - 1}}{\bm{\varphi }}} .
		\end{aligned}
	\end{equation}
	Therefore, the dual problem can be formulated as minimizing ${{\bm{\varphi }}^T}{{{\bf{\tilde V}}}^{ - 1}}{\bm{\varphi }}$, as shown in (\ref{P12}). This completes the proof.

	
	\newpage


\begin{thebibliography}{00}
		\bibitem{mm MIMO}
		S. A. Busari, K. M. S. Huq, S. Mumtaz, L. Dai and J. Rodriguez, ``Millimeter-wave massive MIMO communication for future wireless systems: A survey,'' \textit{IEEE Commun. Surv. Tutor.}, vol. 20, no. 2, pp. 836-869, 2018.
		
		\bibitem{Large MIMO}
		K. Zheng, L. Zhao, J. Mei, B. Shao, W. Xiang, and L. Hanzo, ``Survey of large-scale MIMO systems,'' \textit{IEEE Commun. Surv. Tutor.}, vol. 17, no. 3, pp. 1738–1760, 2015.
		
		\bibitem{MIMO Detection}
		M. A. Albreem, M. Juntti and S. Shahabuddin, ``Massive MIMO detection techniques: A survey,'' \textit{IEEE Commun. Surv. Tutor.}, vol. 21, no. 4, pp. 3109-3132, 2019.
		
		\bibitem{MIMO Linear Precoding}
		M. Vu and A. Paulraj, "MIMO wireless linear precoding," \textit{IEEE Signal Process. Mag.}, vol. 24, no. 5, pp. 86-105, Sept. 2007.
		
		\bibitem{MF}
		R. Esmailzadeh and M. Nakagawa, ``Pre-RAKE diversity combination for direct sequence spread spectrum communications systems,'' in \textit{Proc. IEEE Int. Conf. Commun. (ICC)}, vol. 1, pp. 463–467, May. 1993.
		
		\bibitem{ZF}
		A. Wiesel, Y. C. Eldar and S. Shamai, ``Zero-forcing precoding and generalized inverses,'' \textit{IEEE Trans. Signal Process.}, vol. 56, no. 9, pp. 4409-4418, Sept. 2008.
		
		\bibitem{RZF}
		C. B. Peel, B. M. Hochwald, and A. L. Swindlehurst, ``A vector-perturbation technique for near-capacity multiantenna multiuser communication—Part I: Channel inversion and regularization,'' \textit{IEEE Trans. Commun.}, vol. 53, no. 1, pp. 195–202, Jan. 2005.
		
		\bibitem{SINR balancing}
		M. F. Hanif, L.-N. Tran, A. Tolli, and M. Juntti, ``Computationally efficient robust beamforming for SINR balancing in multicell downlink with applications to large antenna array systems,'' \textit{IEEE Trans. Commun.}, vol. 62, no. 6, pp. 1908–1920, Jun. 2014.
		
		\bibitem{Power Minimization} 
		R. Zhang, S. -H. Leung, H. Wang, W. Tang and Z. Luo, ``Power minimization precoder design for uplink MIMO systems with multi-group NOMA scheme,'' \textit{IEEE Trans. Veh. Technol.}, vol. 70, no. 10, pp. 10553-10569, Oct. 2021.
		
		\bibitem{SDR-star}
		X. Mu, Y. Liu, L. Guo, J. Lin and R. Schober, ``Simultaneously transmitting and reflecting (STAR) RIS aided wireless communications,'' \textit{IEEE Trans. Wireless Commun.}, vol. 21, no. 5, pp. 3083-3098, May. 2022.
		
		\bibitem{SDR-ISAC}
		X. Liu, T. Huang, N. Shlezinger, Y. Liu, J. Zhou and Y. C. Eldar, ``Joint transmit beamforming for multiuser MIMO communications and MIMO radar,'' \textit{IEEE Trans. Signal Process.}, vol. 68, pp. 3929-3944, 2020.
		
		\bibitem{SDR} 
		Z. -q. Luo, W. -k. Ma, A. M. -c. So, Y. Ye and S. Zhang, ``Semidefinite relaxation of quadratic optimization problems'' \textit{IEEE Signal Process. Mag.}, vol. 27, no. 3, pp. 20-34, May. 2010.
		
		\bibitem{Fan-ISAC}
		F. Liu, L. Zhou, C. Masouros, A. Li, W. Luo and A. Petropulu, ``Toward dual-functional radar-communication systems: Optimal waveform design,'' \textit{IEEE Trans. Signal Process.}, vol. 66, no. 16, pp. 4264-4279, 15 Aug. 2018.
		
		\bibitem{Rang-ISAC}
		R. Liu, M. Li, Y. Liu, Q. Wu and Q. Liu, ``Joint transmit waveform and passive beamforming design for RIS-aided DFRC systems,'' \textit{IEEE J. Sel. Topics Signal Process.}, vol. 16, no. 5, pp. 995-1010, Aug. 2022.
		
		\bibitem{Ang CI survey}
		A. Li et al., ``A tutorial on interference exploitation via symbol-level precoding: Overview, state-of-the-art and future directions,'' \textit{IEEE Commun. Surv. Tutor.}, vol. 22, no. 2, pp. 796-839, 2020.
		
		\bibitem{Symeon survey}
		M. Alodeh, D. Spano, A. Kalantari, C. Tsinos, D. Christopoulos, S. Chatzinotas, B. Ottersten, ``Symbol-level and multicast precoding for multiuser multiantenna downlink: A Survey, classification and challenges,'' \textit{IEEE Commun. Surv. Tutor.}, vol. 20, no. 3, pp. 1733-1757, 2018.
		
		\bibitem{CF-PSK}
		A. Li and C. Masouros, ``Interference exploitation precoding made practical: Optimal closed-form solutions for PSK modulations,” \textit{IEEE Trans. Wireless Commun.}, vol. 17, no. 11, pp. 7661–7676, 2018.
		
		\bibitem{CF-QAM}
		A. Li, C. Masouros, B. Vucetic, Y. Li, and A. L. Swindlehurst, ``Interference exploitation precoding for multi-level modulations: Closed-form solutions,” \textit{IEEE Trans. Commun.}, vol. 69, no. 1, pp. 291–308, 2021.
		
		\bibitem{RIRC}
		X. Tong, A. Li, L. Lei, F. Liu and F. Dong, ``Symbol-level precoding for MU-MIMO system with RIRC receiver,'' \textit{IEEE Trans. Commun.}, vol. 72, no. 5, pp. 2820-2834, May. 2024.
		
		\bibitem{SLPMLD}
		X. Tong, A. Li, L. Lei, X. Hu, F. Dong, S. Chatzinotas and C.Masouros, ``MU-MIMO symbol-level precoding for QAM constellations with maximum likelihood receivers,'' 2024, \textit{arXiv:2410.22028}. [Online]. Available: https://arxiv.org/abs/2410.22028.
		
		\bibitem{SLP isac1}
		Y. Zheng, R. Liu, M. Li and Q. Liu, ``End-to-End Learning for SLP-based ISAC Systems,'' in \textit{Proc. IEEE Wireless Commun. Netw. Conf. (WCNC)}, Dubai, United Arab Emirates, 2024.
		
		\bibitem{SLP isac2}
		Z. Liao and F. Liu, `Symbol-level precoding for integrated sensing and communications: A faster-than-Nyquist approach,'' \textit{IEEE Commun. Lett.}, vol. 27, no. 12, pp. 3210-3214, Dec. 2023.
		
		\bibitem{SLP IRS1}
		S. Cai, H. Qu, J. Zhang, X. Shi and H. Zhu, ``Symbol-level precoding design in IRS-aided secure wireless communication systems,'' \textit{IEEE Wireless Commun. Lett.}, vol. 11, no. 11, pp. 2315-2319, Nov. 2022.
		
		\bibitem{SLP IRS2}
		G. Zhang, C. Shen, B. Ai and Z. Zhong, ``Robust symbol-level precoding and passive beamforming for IRS-aided communications,'' \textit{IEEE Trans. Wireless Commun.}, vol. 21, no. 7, pp. 5486-5499, July 2022.
		
		\bibitem{SLP S1}
		Y. Fan, A. Li, X. Liao, and V. C. M. Leung, ``Secure interference exploitation precoding in MISO wiretap channel: Destructive region redefinition with efficient solutions,'' \textit{IEEE Trans. Inf. Forensics Security.}, vol. 16, pp. 402–417, 2021.
		
		\bibitem{SLP S2}
		A. Mayouche, D. Spano, C. G. Tsinos, S. Chatzinotas and B. Ottersten, ``Learning-assisted eavesdropping and symbol-level precoding countermeasures for downlink MU-MISO systems,'' \textit{IEEE Open J. Commun. Soc.}, vol. 1, pp. 535-549, 2020.
		
		\bibitem{Group SLP}
		Z. Xiao, R. Liu, M. Li, Y. Liu, and Q. Liu, ``Low-complexity designs of symbol-level precoding for MU-MISO systems,'' \textit{IEEE Trans. Commun.},vol. 70, no. 7, pp. 4624–4639, Jul. 2022.
		
		\bibitem{BLP CI}
		A. Li, C. Shen, X. Liao, C. Masouros and A. L. Swindlehurst, ``Practical interference exploitation precoding without symbol-by-symbol optimization: A block-level approach,'' \textit{IEEE Trans. Wireless Commun.}, vol. 22, no. 6, pp. 3982-3996, Jun. 2023.
		
		\bibitem{BLP ISAC}
		Y. Wang, X. Hu, A. Li, C. Masouros, K. -K. Wong and K. Yang, ``Symbol-scaling based interference exploitation in ISAC systems: From symbol level to block level,'' \textit{IEEE Trans. Wireless Commun.}, vol. 24, no. 3, pp. 2451-2466, Mar. 2025.
		
		\bibitem{Spano and Alodeh}
		D. Spano, M. Alodeh, S. Chatzinotas and B. Ottersten, ``Symbol-level precoding for the nonlinear multiuser MISO downlink channel,''  \textit{IEEE Trans. Signal Process.}, vol. 66, no. 5, pp. 1331-1345, Mar. 2018.
		
		\bibitem{X ISAC}
		R. Liu, M. Li, Q. Liu and A. L. Swindlehurst, ``Joint symbol-level precoding and reflecting designs for IRS-enhanced MU-MISO systems,'' \textit{IEEE Trans. Wireless Commun.}, vol. 20, no. 2, pp. 798-811, Feb. 2021.
		
		\bibitem{Fast projection}
		L. Condat, ``Fast projection onto the simplex and the $l_1$ ball,'' \textit{Math. Program.}, vol. 158, no. 1/2, pp. 575–585, 2016.
		
		\bibitem{Efficient projection}
		J. Duchi, S. Shalev-Shwartz, Y. Singer, and T. Chandra, ``Efficient projections onto the $l_1$-ball for learning in high dimensions,'' in \textit{ Proc. Int. Conf. Mach. Learn.}, pp. 272–279, 2008.
		
	\end{thebibliography}
\end{document}